\documentclass[aps,pra,twocolumn,groupedaddress]{revtex4-1}

\usepackage{amsmath,amssymb}
\usepackage{graphicx}
\usepackage{hyperref}
\usepackage{txfontsb} 
\usepackage{scrextend}

\hypersetup{
    colorlinks=true,
    linkcolor=blue,
    citecolor=blue,
    filecolor=blue,
    urlcolor=blue,
}

\begin{document}
\title{\vspace*{0.0cm}Assessment of localized and randomized algorithms for electronic structure}
\author{Jonathan E. Moussa}
\altaffiliation[Present address: ]{Molecular Sciences Software Institute, Blacksburg, Virginia 24060, USA}
\email{godotalgorithm@gmail.com}
\author{Andrew D. Baczewski}
\affiliation{Center for Computing Research, Sandia National Laboratories, Albuquerque, New Mexico 87185, USA}

\begin{abstract}
\vspace*{0.55cm}
As electronic structure simulations continue to grow in size, the system-size scaling of computational costs
 increases in importance relative to cost prefactors.
Presently, linear-scaling costs for three-dimensional systems are only attained by localized or randomized algorithms
 that have large cost prefactors in the difficult regime of low-temperature metals.
Using large copper clusters in a minimal-basis semiempirical model as our reference system,
 we study the costs of these algorithms relative to a conventional cubic-scaling algorithm using matrix diagonalization
 and a recent quadratic-scaling algorithm using sparse matrix factorization and rational function approximation.
The linear-scaling algorithms are competitive at the high temperatures relevant for warm dense matter,
 but their cost prefactors are prohibitive near ambient temperatures.
To further reduce costs, we consider hybridized algorithms that combine localized and randomized algorithms.
While simple hybridized algorithms do not improve performance,
 more sophisticated algorithms using recent concepts from structured linear algebra
 show promising initial performance results on a simple-cubic orthogonal tight-binding model.
\vspace*{1.4cm}
\end{abstract}

\maketitle

\section{Introduction\label{intro}}\vspace*{-0.3cm}

The scaling of computational cost with simulation size can significantly impact the usage of a simulation methodology,
 particularly since scientific computing resources have grown exponentially over the last fifty years and will continue to do so for the foreseeable future.
An acute example is molecular dynamics simulations,
 for which atomic forces are calculated by either classical interatomic potential or quantum electronic structure.
From their inception \cite{linear_classical}, interatomic potentials had a linear-scaling cost
 from the locality of repulsion, dispersion, and bonding,
 and sophisticated linear-scaling algorithms were later developed for long-range electrostatic potentials \cite{fast_summation}.
By contrast, early electronic structure computations exhibited the cubic-scaling cost of dense matrix diagonalization \cite{cubic_quantum}.
In both cases, the earliest simulations contained tens of atoms.
Fifty years later, supercomputers can apply interatomic potentials to $10^{12}$ atoms \cite{trillion_classical}
 and conventional electronic structure to $10^{5}$ atoms \cite{hundredthousand_quantum}.
This large size disparity will continue to increase until
 efficient, reliable, and broadly applicable linear-scaling electronic structure algorithms and software are developed.

Significant efforts to develop subcubic-scaling algorithms for electronic structure began in the early 1990's and spanned the decade \cite{linear_quantum}.
This activity focused on localized algorithms that exploit spatial locality of the density matrix, primarily in
 large-gap insulators or at high temperatures where the effect is strongest.
While mean-field calculations were the primary focus, electron correlation calculations with a linear-scaling cost were also established in this period \cite{linear_correlation}.
At a high level, the diverse set of localized algorithms can be categorized by
 whether they work at zero or finite temperature and
 whether they exploit locality in the primary basis or first project the problem into a smaller, localized basis.
These algorithms do not perform well for low-temperature metallic systems,
 with the exception of limited successes for scattering methods \cite{local_scattering} (on free-electron-like materials)
 and energy-renormalization methods \cite{energy_renorm} (when coarse-grained Hamiltonian sparsity can be controlled).
After a decade-long lull in the 2000's, several recent algorithms have reduced scaling without using locality of the density matrix.
Pole expansion and selected inversion (PEXSI) uses mature software for sparse matrix factorization
 to avoid the cubic-scaling bottleneck of dense linear algebra \cite{PEXSI}.
Randomized algorithms use random-vector resolutions of identity to probe density matrices and sample observables \cite{random_quantum}.
With an increasingly diverse set of algorithms, it is ever more difficult to compare their relative cost and accuracy.

A natural performance metric for linear-scaling molecular dynamics algorithms is the cost of an atomic force evaluation per CPU core.
While we are unable to control for differences in computing environments or simulation targets,
 we observe a range of costs over ten orders of magnitude in a survey of recent literature.
AMBER is a popular interatomic potential and software for biomolecular simulations,
 with a typical cost of $\sim$$10^{-6}$ s \cite{AMBER_bench}.
Potentials fit by machine learning are more complicated and expensive,
 with an example cost of $\sim$$10^{-3}$ s \cite{ML_bench}.
Localized density functional theory (DFT) has achieved costs as low as $\sim$$10$ s for large-gap systems \cite{DFT_bench}.
Localized CCSD(T), the ``gold standard'' of quantum chemistry, is more expensive at $\sim$$10^4$ s \cite{CCSDT_bench}.
The large disparity in costs between interatomic potentials and electronic structure can be bridged
 by smaller basis sets and cheaper matrix elements.
Examples of lower-cost electronic structure are
 $\sim$$1$ s for semiempirical quantum chemistry \cite{PM6_bench}
 and $\sim$$10^{-2}$ s for a total-energy tight-binding model \cite{TB_bench}.
New linear-scaling molecular dynamics algorithms should be assessed within this cost spectrum.

In this paper, we assess the cost and accuracy of localized and randomized electronic structure algorithms relative to
 the conventional cubic-scaling algorithm and PEXSI, which has quadratic scaling in three dimensions.
While we cannot test every algorithm, system type, or computing environment,
 we can contribute to a percolation of performance comparisons.
Our software implementation prioritizes shared components between algorithms to enhance simplicity and comparability,
 and source code is available \cite{source_code} for inspection, adaptation, and further benchmarking.
We focus on the difficult case of metallic systems by studying copper clusters of varying size and temperature.
We use a semiempirical tight-binding model in a minimal basis that was fit to reproduce DFT energies \cite{NRL_TB}.
We rationalize and fit benchmark data with cost models that depend on temperature, error, and system size.
This analysis enables clear comparisons between localized and randomized algorithms
 and an assessment of recent attempts to hybridize them \cite{Baer_hybrid,Barros_hybrid}.
This paper expands upon an earlier assessment \cite{random_comment}
 that was focused only on randomized algorithms and did not consider models with realistic materials energetics.

The paper is organized as follows.
In Sec.\@ \ref{methods}, we review the pertinent models and algorithms and summarize our software implementation.
In Sec.\@ \ref{results}, we calibrate costs and errors and present comparative benchmarks.
In Sec.\@ \ref{discussion}, we discuss their implications for future linear-scaling algorithms in electronic structure 
 and prototype two new algorithmic ideas.
In Sec.\@ \ref{conclusions}, we conclude with a reaffirmation and clarification of previous negative assessments
 that available linear-scaling algorithms are still uncompetitive for low-temperature metals.
However, we remain optimistic that there are technical paths forward by
 either developing new low-accuracy algorithms for structured linear algebra
 that combine localization and randomization or
 accepting these approximations as uncontrolled model errors
 to be minimized during the reparameterization of algorithm-specific semiempirical electronic structure models.

\section{Methods\label{methods}}\vspace*{-0.3cm}

Here we provide a focused, self-contained summary of the electronic structure methodology that is relevant to our study.
We abstract away unnecessary details and consolidate several disparate methods into a common theoretical framework and notation.
Bold lowercase and uppercase letters denote vectors and matrices respectively.
All formulae are written implicitly in Hartree atomic units with temperatures written in units of energy ($\hbar = m_e = e = k_e = k_B = 1$).
Some numerical results are written explicitly in units of eV and \AA.

Electronic structure theories with a mean-field form reduce
 the complexity of the many-electron problem to a manageable computational task.
The many-electron Schr\"{o}dinger equation is first projected into a basis of $n$ functions,
 which produces a Hamiltonian matrix with dimension $4^n$.
Mean-field structure exponentially reduces this dimension to $n$ by approximating
 the many-electron ground state using electron orbital vectors
 $\boldsymbol{\phi}_i$ of energy $\epsilon_i$ that satisfy a generalized eigenvalue problem
\begin{equation} \label{eigenproblem}
 \mathbf{H} \boldsymbol{\phi}_i = \epsilon_i \mathbf{S} \boldsymbol{\phi}_i .
\end{equation}
We only consider problems without spin polarization or spin-orbit coupling,
 for which the mean-field Hamiltonian matrix $\mathbf{H}$ and overlap matrix $\mathbf{S}$ are real-valued and symmetric.
When the basis functions are orthogonal, the overlap matrix reduces to the identity matrix, $\mathbf{S} = \mathbf{I}$.
An orthogonalized Hamiltonian, $\mathbf{\overline{H}} \equiv \mathbf{S}^{-1/2} \mathbf{H} \mathbf{S}^{-1/2}$,
 is often used to avoid the complications of nonorthogonal basis functions, but we avoid its use here.

Conventional mean-field calculations are based on solving Eq.\@ (\ref{eigenproblem}),
 but many observables of interest can be defined with $\mathbf{H}$ and $\mathbf{S}$
 rather than $\epsilon_i$ and $\boldsymbol{\phi}_i$.
These observables are based on the free-energy function $g(x)$ and Fermi-Dirac function $f(x)$,
\begin{subequations}
\begin{align}
 g(x) &\equiv -2 T \ln \{ 1 + \exp[-(x-\mu)/T] \} , \label{free_energy_function}\\
 f(x) &\equiv g'(x) = 2 / \{ 1 + \exp[(x-\mu)/T] \} ,
\end{align}
\end{subequations}
 for spin-degenerate electrons at temperature $T$ and chemical potential $\mu$.
The orbital free energy $F$ is a matrix trace over a function of $\mathbf{H}$ and $\mathbf{S}$ that reduces to a sum over functions of $\epsilon_i$,
\begin{equation} \label{free_energy}
 F \equiv \mathrm{tr}[ g( \mathbf{H} \mathbf{S}^{-1} ) ] = \sum_{i=1}^n g(\epsilon_i).
\end{equation}
The linear response of $F$ to changes in $\mathbf{H}$ and $\mathbf{S}$,
\begin{subequations}
\label{density_matrix}
\begin{align}
 \mathbf{P} &\equiv \frac{dF}{d\mathbf{H}} = \mathbf{S}^{-1} f(\mathbf{H} \mathbf{S}^{-1}) = \sum_{i=1}^n f(\epsilon_i) \boldsymbol{\phi}_i \boldsymbol{\phi}_i^T , \\
 \mathbf{Q} &\equiv -\frac{dF}{d\mathbf{S}} = \mathbf{S}^{-1} \mathbf{H} \mathbf{S}^{-1} f(\mathbf{H} \mathbf{S}^{-1}) = \sum_{i=1}^n \epsilon_i f(\epsilon_i) \boldsymbol{\phi}_i \boldsymbol{\phi}_i^T ,
\end{align}
\end{subequations}
 generates the density matrix $\mathbf{P}$ and energy-density matrix $\mathbf{Q}$.
These are the parents of any observables that are based on the linear response of $F$ to changes in an external parameter $\lambda$,
\begin{equation} \label{observable}
 \frac{dF}{d\lambda} = \mathrm{tr}\left[\mathbf{P} \frac{d\mathbf{H}}{d\lambda}\right] - \mathrm{tr}\left[\mathbf{Q} \frac{d\mathbf{S}}{d\lambda}\right] .
\end{equation}
Perhaps the two most important observables are total electron number, $N \equiv \mathrm{tr} [ \mathbf{P} \mathbf{S} ]$, and total orbital energy, $E \equiv \mathrm{tr} [ \mathbf{Q} \mathbf{S} ]$.

For noninteracting electrons at a constant $\mu$, the evaluation
 of $\mathbf{P}$ and $\mathbf{Q}$ for a given $\mathbf{H}$ and $\mathbf{S}$
 can be a complete electronic structure calculation.
However, more realistic calculations of interacting electrons add complications.
The total free energy includes nonlinear dependencies on electronic observables in addition to $F$ that account for electron-electron interactions.
$\mathbf{H}$ also develops a nonlinear observable dependence, the most prevalent being the Hartree potential.
Even for noninteracting electrons, there are nonlinearities of $F$ in external parameters and $\mu$.
The overall effect of these complications is to require multiple calculations of $\mathbf{P}$ and $\mathbf{Q}$ for multiple $\mathbf{H}$ and $\mathbf{S}$,
 so that parameters and observables can be iteratively tuned to satisfy
 physical constraints such as charge neutrality, self-consistent field conditions,
 or atomic relaxation to minimize energy.

Large systems projected into localized basis functions have sparse $\mathbf{H}$ and $\mathbf{S}$,
 which we seek to exploit in reducing the cost of $\mathbf{P}$ and $\mathbf{Q}$.
We express their sparsity with a mask matrix $\mathbf{M}$ that has matrix elements of zero and one corresponding to the
 absence or presence of nonzero matrix elements in $\mathbf{H}$ or $\mathbf{S}$.
If the variation of a parameter $\lambda$ does not change $\mathbf{M}$, then $\mathbf{P}$ and $\mathbf{Q}$ in Eq.\@ (\ref{observable})
 can be replaced by $\mathbf{M} \odot \mathbf{P}$ and $\mathbf{M} \odot \mathbf{Q}$, where $\odot$ denotes the elementwise (Hadamard or Schur) matrix product.
While $\mathbf{P}$ and $\mathbf{Q}$ are not usually sparse,
 the task of computing $\mathbf{M} \odot \mathbf{P}$ and $\mathbf{M} \odot \mathbf{Q}$ from $\mathbf{H}$ and $\mathbf{S}$
 is balanced in the size of its input and output.
Iterative eigensolvers efficiently use sparsity when calculating a small fraction of the $( \epsilon_i , \boldsymbol{\phi}_i )$ eigenpairs,
 but a large fraction contributes to Eq.\@ (\ref{density_matrix}).
Direct eigensolvers are more efficient when calculating all eigenpairs
 but do not use sparsity effectively at their present state of development.
We thus seek to avoid the eigenproblem in Eq.\@ (\ref{eigenproblem}) altogether.

From the many localized linear-scaling electronic structure algorithms that were developed in the 1990's \cite{linear_quantum},
 we consider only Fermi-operator expansion.
Wannier functions are useful for reduced-rank compression of $\mathbf{P}$ and $\mathbf{Q}$ in a large basis set ($n \gg N$),
 but our examples all have small basis sets ($n \approx N$).
Wannier functions induce extra structure in $\mathbf{P}$ and $\mathbf{Q}$ beyond sparsity,
 and we discuss future prospects for structured linear algebra applications to $\mathbf{P}$ and $\mathbf{Q}$ in Sec.\@ \ref{discussion}.
Projector methods were developed for $T=0$, when $\mathbf{P}$ satisfies $\mathbf{P} \mathbf{S} \mathbf{P} = \mathbf{P}$.
While they can approximate a small $T>0$, they are more expensive
 than Fermi-operator expansions for metallic systems \cite{linear_quantum}.

\subsection{NRL tight-binding model}\vspace*{-0.3cm}

To balance between simplicity and utility, we use the Naval Research Laboratory (NRL) tight-binding model \cite{NRL_TB_theory},
 which has a particularly accurate and transferrable fit to copper \cite{NRL_TB}.
Its advantage over other semiempirical total-energy models is the absence of self-consistent field conditions beyond charge neutrality.
Its disadvantage is a restriction to unary materials.
The two most popular semiempirical total-energy models are the modified neglect of differential overlap (MNDO) \cite{MNDO} and
 density-functional tight-binding (DFTB) \cite{DFTB}.
These models all use minimal basis sets of chemically-active atomic orbitals.
MNDO is based on Hartree-Fock theory with Fock exchange restricted to two-center terms.
DFTB is based on DFT with a second-order expansion of electronic charge transfer about an independent-atom reference charge distribution.
The $\mathbf{H}$ and $\mathbf{S}$ produced by these models all have a similar size, spectrum, and sparsity
 (for MNDO, $\mathbf{S} = \mathbf{I}$ and Fock exchange in metals has an unphysical long-range tail that reduces $\mathbf{H}$ sparsity).

\begin{figure}[t!]
\includegraphics{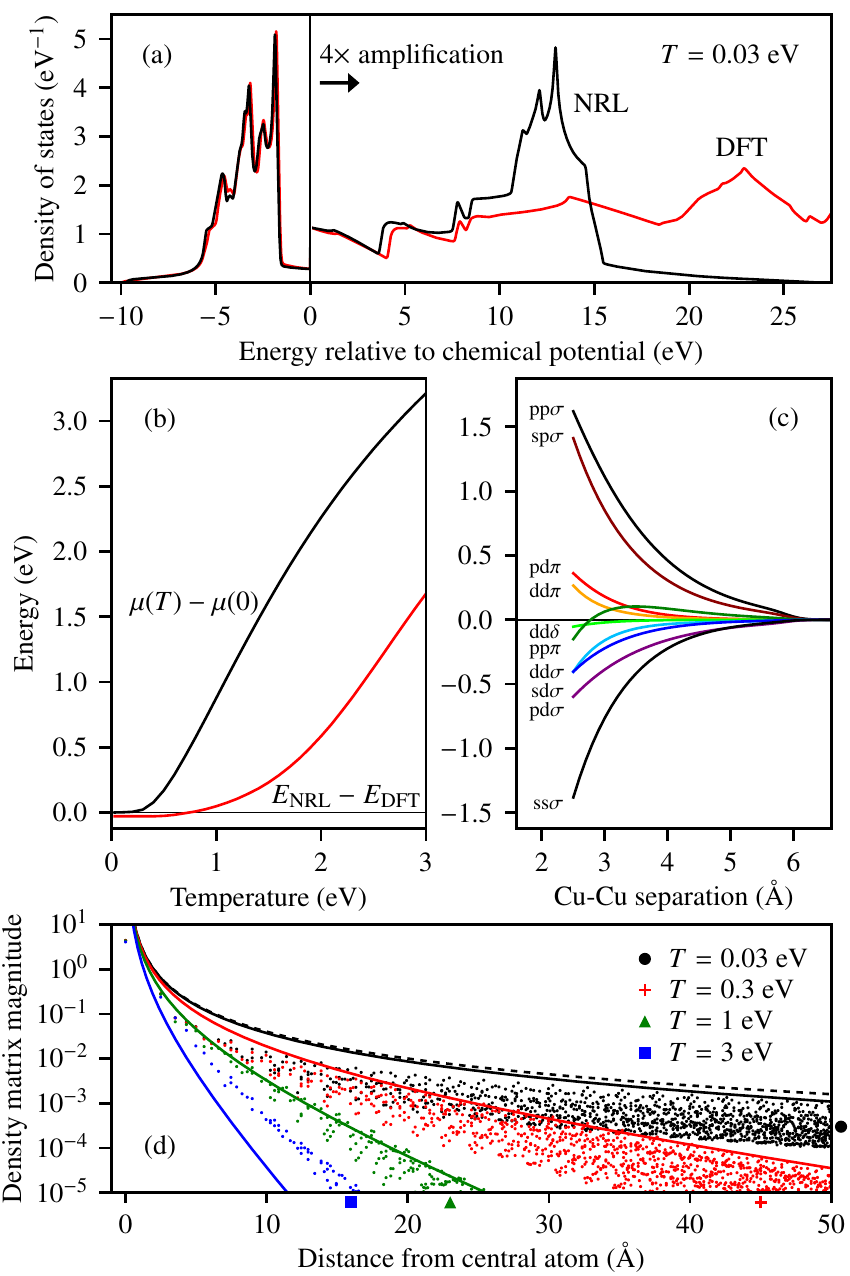}
\caption{\label{NRL_fig} Properties of the NRL tight-binding model for fcc copper, including the (a) electronic density of states,
 (b) chemical potential and per-atom internal energy errors, (c) Slater-Koster hopping matrix elements \cite{Slater_Koster},
 and (d) spatial decay of the density matrix (measured by the Frobenius norm of its $9\times9$ atomic blocks)
 compared with the model in Eq.\@ (\ref{density_decay}) (solid line) and its $T=0$ limit (dashed line).}
\end{figure}

The NRL tight-binding model is effectively an extension of the embedded-atom interatomic potential \cite{EAM} to incorporate electronic structure.
The embedding and pairwise energies of the potential are included in $E$ through the diagonal and off-diagonal matrix elements in $\mathbf{H}$
 for a non-self-consistent tight-binding model of Slater-Koster form \cite{Slater_Koster}.
The total internal and free energies are modeled by $E$ and $F$ with no additional terms.
For copper, the basis includes nine orbitals per atom: 3d, 4s, and 4p.
With an off-diagonal matrix element cutoff of $6.6$ {\AA} and a nearest-neighbor distance of $2.5$ \AA,
 $\mathbf{H}$ and $\mathbf{S}$ for fcc copper have 135 $9\times9$ block matrix elements per block column,
 which takes 87.5 kB of memory in double precision.
Some relevant properties of the NRL Cu model are shown in Fig.\@ \ref{NRL_fig}.
It was fit to DFT data for multiple crystal structures at low $T$
 and transfers up to $T \approx 1$ eV while maintaining errors of less than 0.03 eV/atom in the total energy.
Specifically, the model was fit to DFT internal energies relative to unpolarized reference atoms with the LDA density functional,
 and we use \textsc{vasp} \cite{source_code,VASP} to generate comparable reference data.

For error and cost analysis, we need a simple model for the off-diagonal decay of $\mathbf{P}$ and $\mathbf{Q}$.
We propose a homogenization of the electronic structure problem that is particularly effective for the free-electron-like 4s electrons on the Fermi surface of copper.
First, we use the single-electron Green's function,
\begin{equation}
 \mathbf{G}(\omega) \equiv (\mathbf{H} - \omega \mathbf{S})^{-1},
\end{equation}
 to decompose $\mathbf{P}$ into a Matsubara frequency summation,
\begin{equation}\label{Matsubara}
 \mathbf{P} = \mathbf{S} - 4 T \, \mathrm{Re} \sum_{j = 1}^{\infty} \mathbf{G} [ \mu + (2j-1) \pi T i ] .
\end{equation}
Next, we consider a constant-potential Schr\"{o}dinger operator
 with an effective mass $m^*$ and a potential $\mu - \mu^*$
 as a proxy for $\mathbf{H}$, together with $\mathbf{S} = \mathbf{I}$.
The corresponding proxy for $\mathbf{G}(\omega)$ is a Helmholtz kernel with an exponential decay envelope
\begin{subequations}
\begin{align}
 [\mathbf{G}(\omega)]_{j,k} &\sim \exp(-\gamma |\mathbf{r}_j - \mathbf{r}_k|), \\
 \gamma &= \left| \mathrm{Im} \sqrt{2m^* (\omega + \mu^* - \mu)} \right| ,
\end{align}
\end{subequations}
 where $\mathbf{r}_j$ is the atomic coordinate associated with row $j$.
The exponential decay of $\mathbf{P}$ (and similarly of $g(\mathbf{H}\mathbf{S}^{-1})$ and $\mathbf{Q}$)
 is set by the $\mathbf{G}(\omega)$ in Eq.\@ (\ref{Matsubara}) with the smallest decay exponent,
\begin{equation}
 \gamma_{\min} = \left| \mathrm{Im} \sqrt{2m^* ( \mu^* + \pi T i )} \right| .
\end{equation}
In this homogeneous model, metals correspond to $\mu^* > 0$ and insulators correspond to $\mu^* < 0$.
We could either use $\mu^*$ as a tuning parameter or fit it to a model.
For free electrons with density $\rho$, $\mu^* = ( 3 \pi^2 \rho )^{2/3} / (2 m^*) $.
For insulators with electron-hole symmetry, $\mu^*$ is half of the band gap.
These results are consistent with existing models of density-matrix decay \cite{DM_decay}
 while fully specifying $\gamma_{\min}$.
The same $\gamma_{\min}$ applies to $\mathbf{Q}$ and $g(\mathbf{HS}^{-1})$
 because like $\mathbf{P}$ they are both holomorphic functions
 of $\mathbf{H}\mathbf{S}^{-1}$ outside of $\mu +  \xi i$ for $\xi \in [-\infty,-\pi T] \cup [\pi T,\infty]$.

The exponential off-diagonal decay is weak enough for Cu at ambient temperature
 that algebraic decay must be added to the model.
For simplicity, we use the asymptotic power laws at $T=0$ established in a recent study \cite{Barros_hybrid}.
We build a crude interpolation between the $T=0$ and high-$T$ limits by simply
 multiplying the exponential and algebraic decay envelopes,
\begin{subequations} \label{decay_model}
\begin{align}
 [\mathbf{G}(\mu \pm \pi T i)]_{j,k} & \sim  \frac{\exp(-\gamma_{\min} |\mathbf{r}_j - \mathbf{r}_k| ) }{|\mathbf{r}_j - \mathbf{r}_k|^{(D-1)/2}}, \\
 [\mathbf{P}]_{j,k} & \sim \frac{\exp(-\gamma_{\min} |\mathbf{r}_j - \mathbf{r}_k| ) }{|\mathbf{r}_j - \mathbf{r}_k|^{(D+1)/2}} , \label{density_decay} \\
  [ g(\mathbf{H}\mathbf{S}^{-1}) ]_{j,k} \sim [\mathbf{Q}]_{j,k} & \sim \frac{\exp(-\gamma_{\min} |\mathbf{r}_j - \mathbf{r}_k| ) }{|\mathbf{r}_j - \mathbf{r}_k|^{(D+3)/2}},
\end{align}
\end{subequations}
 for homogeneous systems in $D$ spatial dimensions.
We fit the model envelope in Fig.\@ \ref{NRL_fig} with $m^* = 1$ and $\mu^* = 10$ eV,
 while the free-electron value is 7.4 eV assuming that only the 4s Cu electrons contribute to the metallic state.
This model is only intended to be used as a simple heuristic, and there are many other studies \cite{decay_review}
 of off-diagonal operator decay in electronic structure with varying amounts of mathematical rigor.

\subsection{Function approximations}\vspace*{-0.3cm}

Efficient evaluation of a matrix function usually requires it to be approximated by more operationally convenient matrix functions.
Polynomials, rational functions, and exponentials are important examples, in decreasing order of convenience.
Matrix polynomials can be constructed recursively through a sequence of matrix-matrix or matrix-vector operations.
Pole decompositions of matrix rational functions can be evaluated by inverting matrices or solving the associated linear systems.
Matrix exponentials can be evaluated by numerical integration of linear differential equations.
Electronic structure methods sometimes use additional intermediate functions to facilitate novel algorithms.
For example, multi-$T$ telescoping series \cite{energy_renorm} approximate sharper features over narrower intervals,
 product expansions \cite{random_path} enable randomized determinant calculations, 
  and temperature-halving transformations \cite{ONETEP_metal} express low-$T$
  density matrices by nesting higher-$T$ density matrices.
Here, we only utilize direct polynomial and rational approximations of $f(x)$ and discuss extensions to $g(x)$.

\subsubsection{Polynomial approximation}\vspace*{-0.3cm}

Following standard practice, we construct our polynomial approximations using Chebyshev polynomials
 after mapping the approximation domain to $[-1,1]$.
For a spectral interval $[\epsilon_{\min},\epsilon_{\max}]$, we fit $\hat{f}(x) \equiv f [ \epsilon_{\min} (1 - x) / 2 + \epsilon_{\max} (1 + x) / 2 ]$
 to $p$ Chebyshev polynomials $C_k(\cos \theta) \equiv \cos( k \theta )$ with a projection
 using the inner product under which they are orthogonal,
\begin{subequations}
\begin{align}
 \hat{f}(x) &\approx \sum_{k=0}^{p-1} \alpha_k C_k(x), \\
 \alpha_k &= \frac{2 - \delta_{k,0}}{\pi} \int_{-1}^{1} C_k(x) \hat{f}(x) \frac{dx}{\sqrt{1 - x^2}} .
\end{align}
\end{subequations}
Utilizing Chebyshev-Gauss quadrature and a discrete Fourier transform, we efficiently and accurately calculate $\alpha_k$.
We can model the observed maximum pointwise error in $\hat{f}(x)$ as
\begin{equation}
 \varepsilon_{\mathrm{poly}} \approx \exp \left( - p \frac{6.4}{(\epsilon_{\max} - \epsilon_{\min})/T} \right)  .
\end{equation}
For the NRL Cu model, $\epsilon_{\max} - \epsilon_{\min} \approx 42$ eV.
Direct minimax optimization can further reduce $\varepsilon_{\mathrm{poly}}$, but its scaling does not change.
The same approximation process applies to $g(x)$.

We use the polynomial approximation of $\hat{f}(x)$ to calculate $\mathbf{P}\mathbf{y}$ and $\mathbf{Q}\mathbf{y}$ for a given vector $\mathbf{y}$.
With a mapped Hamiltonian, $\hat{\mathbf{H}} \equiv [2 \mathbf{H} - (\epsilon_{\max} + \epsilon_{\min}) \mathbf{S}] /(\epsilon_{\max} - \epsilon_{\min})$,
 we use Chebyshev recursion relations to evaluate $\mathbf{y}_k \equiv C_k( \hat{\mathbf{H}} \mathbf{S}^{-1} )\mathbf{y}$,
\begin{subequations} \label{poly_PQ}
\begin{align}
 \left[ \begin{array}{c} \mathbf{P}\mathbf{y} \\ \mathbf{Q}\mathbf{y} \end{array} \right] &\approx
  \left[ \begin{array}{c} \mathbf{S}^{-1} \\ \mathbf{S}^{-1} \mathbf{H} \mathbf{S}^{-1} \end{array} \right] \sum_{k=0}^{p-1} \alpha_k \mathbf{y}_k , \\
 \mathbf{y}_0 &= \mathbf{y}, \\
 \mathbf{y}_1 &= \hat{\mathbf{H}} \mathbf{S}^{-1} \mathbf{y}_0 , \\
 \mathbf{y}_{k+1} &=  2 \hat{\mathbf{H}} \mathbf{S}^{-1} \mathbf{y}_k - \mathbf{y}_{k-1} .
\end{align}
\end{subequations}
This requires intermediate calculations of $\mathbf{t}_i \equiv \mathbf{S}^{-1}\mathbf{y}_i$.
We can either precompute and apply a sparse approximation of $\mathbf{S}^{-1}$ or
 iteratively solve sparse linear systems, $\mathbf{S} \mathbf{t}_i = \mathbf{y}_i$.
We choose to solve the linear systems by using the conjugate gradient (CG) method
 and examine the alternative method in Sec.\@ \ref{preconditioners}.

\subsubsection{Rational approximation}\vspace*{-0.3cm}

Because of the poles close to the real axis at $\omega = \mu \pm \pi T i$ in Eq.\@ (\ref{Matsubara}),
 $f(x)$ is more efficiently approximated with a rational function than a polynomial.
There are both analytical \cite{analytical_rational_fit} and numerical \cite{numerical_rational_fit} approximations available.
The best analytical approximation can be applied to either $f(x)$ or $g(x)$ assuming a finite spectral interval $[\epsilon_{\min},\epsilon_{\max}]$ as in the polynomial case.
The best numerical approximation relies on an unstable fitting process that has only been applied to $f(x)$
 but uses four times fewer poles and depends on $\mu - \epsilon_{\min}$ instead of $\epsilon_{\max} - \epsilon_{\min}$.
A numerical fit of $f(x)$ with $p$ pole pairs in residue-pole form,
\begin{equation} \label{rational_fit}
 f(x) \approx 2 \, \mathrm{Re} \left[ \sum_{k=1}^{p} \frac{w_k}{x - z_k} \right] ,
\end{equation}
 has an empirical pointwise maximum error of 
\begin{equation}
 \varepsilon_{\mathrm{rational}} \approx 4 \exp \left( - p \frac{9.9}{\ln[ 3.1 (\mu - \epsilon_{\min})/T]} \right)
\end{equation}
 for $T \le 0.1 (\mu - \epsilon_{\min})$.
For a relevant example of $T = 0.03$ eV and $\varepsilon \le 10^{-3}$,
 we require $p=1470$ for the polynomial fit but only $p=6$ for the rational function fit.

The efficiency of the rational approximation is partly offset by an increased amount of work per fitting function.
First, we order the poles by their distance from $\mu$, $| z_{k+1} - \mu | < | z_k - \mu |$.
We then decompose $\mathbf{P} \mathbf{y}$ and $\mathbf{Q} \mathbf{y}$ for a given real $\mathbf{y}$ into linear systems
 that determine the intermediate vectors $\mathbf{g}_k \equiv \mathbf{G}(z_k) \mathbf{y}$,
\begin{subequations} \label{rational_PQ}
\begin{align}
 \mathbf{P}\mathbf{y} &= 2 \, \mathrm{Re} \left[ \sum_{k=1}^p w_k \mathbf{g}_k \right] , \\
 \mathbf{Q}\mathbf{y} &= 2 \, \mathrm{Re} \left[ \mathbf{S}^{-1} \mathbf{y} \sum_{k=1}^p w_k + \sum_{k=1}^p w_k z_k \mathbf{g}_k \right] , \\
 (\mathbf{H} - z_k \mathbf{S}) \mathbf{g}_k &= \mathbf{y} .
\end{align}
\end{subequations}
The condition number $\kappa$ of these linear systems increases as the $z_k$ approach the real axis at $\mu$
 with a $T$-dependent bound,
\begin{align}
 \kappa  & \approx \frac{\max\{ | \epsilon_{\max} - \mu | , | \epsilon_{\min} - \mu |\}}{|\mathrm{Im}(z_k)|} \notag \\
  & > \frac{\max\{ | \epsilon_{\max} - \mu | , | \epsilon_{\min} - \mu |\}}{\pi T} . \label{condition_number0}
\end{align}
If we solve for $\mathbf{g}_k$ in sequential order, we can initialize $\mathbf{g}_{k+1}$ to $\mathbf{g}_k$
 in an iterative linear solver and precondition it with some approximation of $\mathbf{G}(z_k)$
 that was derived from $\mathbf{g}_k$ for multiple $\mathbf{y}$.
We apply the CGLS iterative linear solver \cite{CGLS}
 to maintain the simple structure of the CG method for indefinite $\mathbf{H} - z_k \mathbf{S}$ matrices.
Without an accurate preconditioner, a large number of solver iterations will be required for small $|\mathrm{Im}(z_k)|$.
The use of various preconditioners is examined in Sec.\@ \ref{preconditioners}.

\subsection{Trace approximations\label{trace_approx}}\vspace*{-0.3cm}

We consider approximations of the matrix traces in Eq.\@ (\ref{observable})
 that are based on approximations of $\mathbf{M}\odot\mathbf{P}$ and $\mathbf{M}\odot\mathbf{Q}$.
Such approximations are generically denoted as $\mathbf{M}\odot\tilde{\mathbf{P}}$ and $\mathbf{M}\odot\tilde{\mathbf{Q}}$,
 and they produce approximate observables $d\tilde{F}/d\lambda$.
With this structure, we can bound all errors in Eq.\@ (\ref{observable}) as
\begin{align} \label{error_bound}
 \left| \frac{dF}{d\lambda} - \frac{d\tilde{F}}{d\lambda} \right| &\le \left\| \frac{d\mathbf{H}}{d\lambda} \right\|_F \left\| \mathbf{M}\odot(\mathbf{P} - \tilde{\mathbf{P}}) \right\|_F \notag \\
  & \ \ \ \ + \left\| \frac{d\mathbf{S}}{d\lambda} \right\|_F \left\| \mathbf{M}\odot(\mathbf{Q} - \tilde{\mathbf{Q}}) \right\|_F
\end{align}
using the triangle and Cauchy-Schwartz inequalities with the Frobenius norm, $\| \mathbf{X} \|_F \equiv \! \sqrt{\mathrm{tr}(\mathbf{X}^\dag \mathbf{X})}$.
While every observable may have its own practical accuracy target, this analysis can relate them to a common target for $\mathbf{M}\odot\mathbf{P}$ and $\mathbf{M}\odot\mathbf{Q}$.
This is not a tight error bound in practice, but it is a useful reference
 point from which to discuss more accurate error estimates for
 special cases with additional matrix structure to exploit.

For both exact and approximate algorithms, calculations of $\mathbf{M}\odot\mathbf{P}$ and $\mathbf{M}\odot\mathbf{Q}$ can be reduced to numerical linear algebra primitives.
The generalized eigenvalue problem in Eq.\@ (\ref{eigenproblem}) is the standard primitive.
A large fraction of its eigenvalues and eigenvectors contribute to Eq.\@ (\ref{density_matrix}) in small-basis calculations.
In this regime, there are no known algorithms that efficiently exploit matrix sparsity,
 and dense-matrix algorithms with an $O(n^3)$ cost are commonly used.
LAPACK \cite{LAPACK} is the standard implementation for shared-memory computations.
To exploit matrix sparsity, approximations are necessary.
For example, matrix functions can be replaced by rational approximations as in Eq.\@ (\ref{rational_fit}),
 which decomposes the calculation of $\mathbf{M}\odot\mathbf{P}$ or $\mathbf{M}\odot\mathbf{Q}$ into a sum over $\mathbf{M}\odot\mathbf{G}(z_i)$.
Selected matrix inversion as implemented in PEXSI \cite{PEXSI} is able to calculate the matrix elements of $\mathbf{A}^{-1}$
 in the sparsity pattern of $\mathbf{A}$ from a sparse LU decomposition of $\mathbf{A}$
 with a cost similar to the decomposition itself.
In a nested-dissection ordering, LU decomposition of a sparse matrix with local connectivity
 in $D$ spatial dimensions has a canonical $O(n^{3-\min\{3/D,2\}})$ cost.
While further reductions to an $O(n)$ cost require more complicated approximations, the
 examples that follow are all based on the $\mathbf{P}\mathbf{y}$ and $\mathbf{Q}\mathbf{y}$ matrix-vector primitives
 discussed in the previous subsection.

\subsubsection{Local approximation\label{local_approx}}\vspace*{-0.3cm}

Localized linear-scaling calculations of $\mathbf{M}\odot\mathbf{P}$ and $\mathbf{M}\odot\mathbf{Q}$
 assume that all intermediate steps are confined to a restricted sparsity pattern.
We define this pattern with a localized mask matrix $\mathbf{M}_L(r_{\max})$ that depends on a localization radius $r_{\max}$,
\begin{equation}
 [\mathbf{M}_L(r)]_{j,k} = \left\{ \begin{array}{cc} 1 , & |\mathbf{r}_j - \mathbf{r}_k| \le r \\ 0 , & |\mathbf{r}_j - \mathbf{r}_k| > r \end{array} \right. .
\end{equation}
Matrix elements of the NRL Cu model have a cutoff radius $r_0$, therefore $\mathbf{M} = \mathbf{M}_L(r_0)$.
In general, the sparsity pattern may be defined using additional information and dynamically adapted to minimize observed localization errors.
To spatially restrict matrix operations, we define local matrices $\mathbf{L}_i$ that are the $n_i$ columns of $\mathbf{I}$
 corresponding to the nonzero columns in the $i$th row of $\mathbf{M}_L(r_{\max})$.
They define $n$ local restrictions of $\mathbf{H}$ and $\mathbf{S}$,
\begin{equation}
 \mathbf{H}_i \equiv \mathbf{L}_i^T \mathbf{H} \mathbf{L}_i \ \ \ \mathrm{and} \ \ \ \mathbf{S}_i \equiv \mathbf{L}_i^T \mathbf{S} \mathbf{L}_i ,
\end{equation}
 with matrix dimensions $n_i$ that depend on $r$ but not $n$.
A local calculation will use an $\mathbf{H}_i$ and $\mathbf{S}_i$ rather than $\mathbf{H}$ and $\mathbf{S}$.

The local approximation is based on an incomplete, vector-dependent resolution of identity.
For each natural basis vector $\mathbf{e}_i$, all matrix operations applied to it are projected by a local
 resolution of identity $\mathbf{L}_i \mathbf{L}_i^T$ that induces the approximation
\begin{equation}\label{local_projection}
 \left[ \prod_{j=1}^m \mathbf{X}_j \right] \mathbf{e}_i \approx \mathbf{L}_i \left[ \prod_{j=1}^m (\mathbf{L}_i^T \mathbf{X}_j \mathbf{L}_i) \right] \mathbf{L}_i^T \mathbf{e}_i .
\end{equation}
This is exact when a product of matrices $\mathbf{X}_i$ and a sequence of intermediate products are contained inside the sparsity pattern of $\mathbf{M}_L(r)$,
 which in practice only holds for small $m$.
The local matrix projections of $\mathbf{P}$ and $\mathbf{Q}$ induced by Eq.\@ (\ref{local_projection}) are
\begin{equation}
 \mathbf{P}_i \equiv \mathbf{S}_i^{-1} f(\mathbf{H}_i \mathbf{S}_i^{-1})  \ \ \ \mathrm{and} \ \ \ \mathbf{Q}_i \equiv \mathbf{S}_i^{-1} \mathbf{H}_i \mathbf{S}_i^{-1} f(\mathbf{H}_i \mathbf{S}_i^{-1}).
\end{equation}
The local approximations of $\mathbf{M}\odot\mathbf{P}$ and $\mathbf{M}\odot\mathbf{Q}$ are then
\begin{subequations} \label{local_density}
\begin{align}
 \mathbf{M}\odot\tilde{\mathbf{P}}_L &\equiv \sum_{i=1}^n \mathbf{L}_i \mathbf{P}_i \mathbf{L}_i^T \mathbf{e}_i \mathbf{e}_i^T, \\
 \mathbf{M}\odot\tilde{\mathbf{Q}}_L &\equiv \sum_{i=1}^n \mathbf{L}_i \mathbf{Q}_i \mathbf{L}_i^T \mathbf{e}_i \mathbf{e}_i^T  .
\end{align}
\end{subequations}
Local matrices are symmetric, but they break exact symmetry in $\mathbf{M}\odot\tilde{\mathbf{P}}_L$ and $\mathbf{M}\odot\tilde{\mathbf{Q}}_L$
 by contributing to a single column.

We apply the homogeneous spatial decay model in Eq.\@ (\ref{decay_model}) to estimate localization errors in Eq.\@ (\ref{error_bound}).
We assume that the effect of a local projection is to enforce a zero-value boundary
 condition at $|\mathbf{r}_j - \mathbf{r}_k| = r_{\max}$ for a calculation centered at $\mathbf{r}_j$,
\begin{align} \label{homogeneous_local_density}
 [\tilde{\mathbf{P}}_L]_{j,k} &\sim \left[ \frac{\exp(-\gamma_{\min} |\mathbf{r}_j - \mathbf{r}_k| ) }{|\mathbf{r}_j - \mathbf{r}_k|^{(D+1)/2}} - \frac{\exp(-\gamma_{\min} r_{\max} ) }{r_{\max}^{(D+1)/2}} \right] \notag \\
 & \ \ \ \ \times \theta( r_{\max} - |\mathbf{r}_j - \mathbf{r}_k| ) ,
\end{align}
 where $\theta(x)$ is the Heaviside step function.
To homogenize the Frobenius norm, we replace the trace with an integral where
 $r_{\min}$ defines an average volume per basis function,
\begin{align} \label{local_homogeneous_bound}
 \left\| \mathbf{M}\odot(\mathbf{P} - \tilde{\mathbf{P}}_L) \right\|_F &\sim \sqrt{ n \int_{0}^{r_0} \frac{\exp(-2\gamma_{\min} r_{\max} ) }{r_{\max}^{D+1}} \frac{r^{D-1} dr}{r_{\min}^D} } \notag \\
  &\sim \frac{\exp(-\gamma_{\min} r_{\max} )}{r_{\max}^{(D+1)/2}} \sqrt{ \frac{n r_0^D}{r_{\min}^D} } .
\end{align}
The same analysis applies to $\tilde{\mathbf{Q}}_L$, which decays more rapidly with an additional factor of $r_{\max}^{-1}$.
Thus an off-diagonal spatial decay of density matrix elements is directly proportional to a reduction of observable error bounds with increasing $r_{\max}$.

The local calculations in Eq.\@ (\ref{local_density}) involve $n$ distinct matrix problems of size $n_i \sim r_{\max}^D$ in $D$ spatial dimensions.
The total cost of these calculations is $\sim m r_{\max}^D n$, where $m$ is the average number of matrix-vector multiplications required to calculate $f(\mathbf{H}_i \mathbf{S}_i^{-1}) \mathbf{y}$.
We ignore the $n$-dependence of errors in Eq.\@ (\ref{local_homogeneous_bound}) to characterize the local error per atom $\varepsilon_{\mathrm{local}}$.
The relationship between $r_{\max}$, $T$, and $\varepsilon_{\mathrm{local}}$ in the homogeneous model is
\begin{equation} \label{local_radius}
 r_{\max} \sim \left\{ \begin{array}{lr} \left(\frac{1}{\varepsilon_{\mathrm{local}}}\right)^{2/(D+1)} - \frac{T}{\sqrt{\mu^*}} \left(\frac{1}{\varepsilon_{\mathrm{local}}}\right)^{4/(D+1)}, & T \ll \mu^* \\
  \frac{1}{\sqrt{T}} \ln\left( \frac{1}{\varepsilon_{\mathrm{local}}} \right) , & T \gg \mu^* \end{array} \right. .
\end{equation}
The model clearly articulates the significant difference in cost between low-$T$ and high-$T$ regimes.
In the low-$T$ limit, $r_{\max}$ increases algebraically with increasing accuracy and $T$ causes a sub-leading-order effect.
In the high-$T$ limit, $r_{\max}$ increases logarithmically with increasing accuracy and $T$ has a leading-order effect,
 similar to insulators and their energy gap.

\subsubsection{Random approximation}\vspace*{-0.3cm}

Randomized linear-scaling calculations are based on direct evaluations of $\mathbf{P} \mathbf{y}$ and $\mathbf{Q} \mathbf{y}$ for multiple random vectors $\mathbf{y}$
 from which $\mathbf{M}\odot\mathbf{P}$ and $\mathbf{M}\odot\mathbf{Q}$ are approximated.
Pseudorandom vectors forming columns of matrices $\mathbf{R}_i$ are used to construct random projections $\mathbf{R}_i \mathbf{R}_i^\dag$,
 which preserve distances between vectors according to the Johnson-Lindenstrauss lemma \cite{random_projection}.
For this application, the mean value of $\mathbf{R}_i \mathbf{R}_i^\dag$ should be $\mathbf{I}$,
\begin{equation} \label{random_RoI}
  \lim_{s \rightarrow \infty} \frac{1}{s} \sum_{i=1}^s \mathbf{R}_i \mathbf{R}_i^\dag = \mathbf{I},
\end{equation}
 and its elementwise standard deviation matrix,
\begin{equation}
   [\mathbf{E}]_{j,k} \equiv \sqrt{\lim_{s \rightarrow \infty} \frac{1}{s} \sum_{i=1}^s | [ \mathbf{R}_i \mathbf{R}_i^\dag - \mathbf{I} ]_{j,k}|^2},
\end{equation}
 is necessary to estimate finite-sampling errors.
An example is the random-phase vector ensemble,
 where $\mathbf{R}_i$ has one column and its matrix elements are complex with unit amplitude and uniformly random phase.
Its elementwise standard deviation is $[\mathbf{E}]_{j,k} = 1 - \delta_{j,k}$.
The primary approximation of randomized methods is a replacement of $\mathbf{I}$ with averages over $s$ instances of $\mathbf{R}_i \mathbf{R}_i^\dag$,
 which incurs an $O(s^{-1/2})$ finite-sampling error.

The random approximation, like the local approximation, is effectively an incomplete resolution of identity.
However, the random resolution of identity is equally valid for any vector, 
 whereas a local resolution of identity is adapted to a specific $\mathbf{e}_i$.
The generic random approximations of $\mathbf{X}$ and $\mathbf{M}\odot\mathbf{X}$ are
\begin{subequations} \label{random_density}
\begin{align}
 \tilde{\mathbf{X}}_R &\equiv \frac{1}{s} \sum_{i=1}^s   \mathbf{X} \mathbf{R}_i \mathbf{R}_i^\dag , \\
 \mathbf{M}\odot\tilde{\mathbf{X}}_R &\equiv \frac{1}{s} \sum_{i=1}^s  \mathbf{M}\odot \left( \mathbf{X} \mathbf{R}_i \mathbf{R}_i^\dag \right) ,
\end{align}
\end{subequations}
analogous to Eq.\@ (\ref{local_density}).
We are able to probe $\mathbf{P}$ and $\mathbf{Q}$ directly, rather than their local subspace projections.
Because Eq.\@ (\ref{random_density}) is linear in $\mathbf{R}_i \mathbf{R}_i^\dag$, it samples from $\mathbf{P}$ and $\mathbf{Q}$ without bias.
With finite sampling errors, the real-symmetric $\mathbf{M}\odot\mathbf{P}$ and $\mathbf{M}\odot\mathbf{Q}$
 relax to complex, nonsymmetric $\mathbf{M}\odot\tilde{\mathbf{P}}_R$ and $\mathbf{M}\odot\tilde{\mathbf{Q}}_R$.

Random approximations are more amenable to estimates of typical errors rather than strict error bounds,
 but we can still relate them to the observable error bounds in Eq.\@ (\ref{error_bound}).
For a large number of samples, matrix trace errors should obey the central limit theorem with zero mean and standard errors of
\begin{equation}\label{sampling_error}
 \left| \mathrm{tr}\left[\mathbf{X} - \tilde{\mathbf{X}}_R \right] \right| \approx \frac{\| \mathbf{X} \odot \mathbf{E} \|_F}{\sqrt{s}} .
\end{equation}
We can bound this estimate with a similar form as Eq.\@ (\ref{error_bound}) by
 splitting $\mathbf{X}$ into sparse $\mathbf{A} = \mathbf{M} \odot \mathbf{A}$ and dense $\mathbf{Y}$ and bounding
 each matrix separately by maximizing over sparse matrices,
\begin{align} \label{estimate_bound}
 \| (\mathbf{A} \mathbf{Y}) \odot \mathbf{E} \|_F &\le \| \mathbf{A} \|_F \max_{\mathbf{B} = \mathbf{M} \odot \mathbf{B}}\frac{\| (\mathbf{B} \mathbf{Y}) \odot \mathbf{E} \|_F}{\| \mathbf{B} \|_F} \notag \\
 &= \| \mathbf{A} \|_F \max_i \| \mathbf{L}_i^T \mathbf{Y} \mathbf{D}_i \|_2 ,
\end{align}
 where $[\mathbf{D}_i]_{j,k} \equiv [\mathbf{E}]_{i,j} \delta_{j,k}$ and $\mathbf{L}_i$ are the local matrices defined by $\mathbf{M}$ in Sec.\@ \ref{local_approx}.
Alternatively, we can estimate the bound in Eq.\@ (\ref{error_bound}) for a split $\mathbf{X} = \mathbf{A} \mathbf{Y}$, which has a mean value of
\begin{align} \label{bound_estimate}
 \| \mathbf{A} \|_F \, \| \mathbf{M} \odot (\mathbf{Y} - \tilde{\mathbf{Y}}_R) \|_F &\approx \| \mathbf{A} \|_F \frac{ \| \mathbf{M} \odot (\mathbf{Y} \mathbf{E}) \|_F}{\sqrt{s}} \notag \\
  &= \| \mathbf{A} \|_F \sqrt{ \frac{1}{s} \sum_{i=1}^n \| \mathbf{L}_i^T \mathbf{Y} \mathbf{D}_i \|_F^2 }.
\end{align}
The estimated bound is looser than the bounded estimate by a factor of $\approx \! n^{1/2}$,
 exemplifying the benefits that a typical error analysis can have over a worst-case error analysis.

We again use the homogeneous model to construct random error estimates comparable to Eq.\@ (\ref{local_homogeneous_bound}).
For this purpose, we use the error estimate in Eq.\@ (\ref{bound_estimate}),
 because the tighter estimate in Eq.\@ (\ref{estimate_bound}) does not have an available local-error analog.
For the random-phase ensemble, it is the matrix product between $\mathbf{Y}$ and $\mathbf{E}$ in Eq.\@ (\ref{bound_estimate}) that we replace with a spatial integral,
\begin{align} \label{random_homogeneous_bound}
 \left\| \mathbf{M}\odot(\mathbf{P} - \tilde{\mathbf{P}}_R) \right\|_F &\sim \sqrt{ \frac{n r_0^D}{s r_{\min}^D} \int_{r_{\min}}^{\infty} \frac{\exp(-2\gamma_{\min} r ) }{r^{D+1}} \frac{r^{D-1} dr}{r_{\min}^D} } \notag \\
  &\sim \frac{\exp(-\gamma_{\min} r_{\min} )}{r_{\min}^{(D+1)/2}} \sqrt{ \frac{n r_0^D}{s r_{\min}^D} } ,
\end{align}
 with the same result for $\tilde{\mathbf{Q}}_R$.
Random errors are insensitive to the presence or absence of spatial decay in the density matrix,
 in agreement with previous results \cite{random_quantum}.
However, previously observed self-averaging of random errors is absent from this analysis.
In Sec.\@ \ref{self_averaging}, we show that self-averaging errors
 are not generic but a special property of specific observables.

The random calculations in Eq.\@ (\ref{random_density}) are based on matrices of size $n$,
 but they are limited in quantity to $s$ times the number of columns in $\mathbf{R}_i$.
This small number of large matrix problems is a complementary distribution of computational work to the
 large number of small matrix problems in local calculations.
The random error per basis function in Eq.\@ (\ref{random_homogeneous_bound}) is set by the
 standard finite-sampling error, $\varepsilon_{\mathrm{random}} \sim s^{-1/2}$.
Thus the error cost prefactor of random calculations is $\varepsilon_{\mathrm{random}}^{-2}$,
 compared to $\varepsilon_{\mathrm{local}}^{-2+2/(D+1)}$ for low-$T$ local calculations according to Eq.\@ (\ref{local_radius}).
This analysis suggests that increasing accuracy requirements will cause local calculations to be more efficient than random calculations,
 especially at high temperatures or for insulators.
However, this outcome depends on $\mathbf{R}_i$, and there are efficient alternatives to the random-phase vector ensemble \cite{Barros_hybrid}.

\subsubsection{Hybrid approximation\label{hybrid_approx}}\vspace*{-0.3cm}

The simple rationale for hybrid methods is that localization and randomization are complementary approximations
 where the primary limitation of each method can be repaired by the other.
Local calculations can reduce the sampling variance of random calculations
 by providing localized approximations of the density matrix
 to enable sampling from the residual error in the density matrix rather than the full density matrix itself.
Random calculations can sample from the full density matrix to construct models of the electronic environment
 in which to embed local calculations and reduce localization errors.

Two hybrid local-random linear-scaling electronic structure algorithms have been demonstrated.
The first demonstration \cite{Baer_hybrid} constructs a local approximation by divide-and-conquer
 decomposition of nanostructures into independent molecular fragments.
The random component of this algorithm uses the random-phase vector ensemble,
 where the full density matrix contributes to the sampling variance and the fragment density matrices provide significant variance reduction.
The effective second demonstration \cite{Barros_hybrid} does not include any explicit local calculations.
Instead, localization is introduced with a multi-color vector ensemble that partitions the basis functions into $q$ colors,
 where $c(i)$ is the color of the $i$th function satisfying $|\mathbf{r}_i - \mathbf{r}_j| > r_{\mathrm{max}}$ whenever $c(i) = c(j)$.
The associated $\mathbf{R}_i$ has $q$ columns with complex random-phase elements satisfying
\begin{subequations}
\begin{align}
 | [ \mathbf{R}_i ]_{j,k} | &= \left\{ \begin{array}{ll} 1, & c(j) = k \\ 0 , & c(j) \neq k \end{array} \right. , \\
 [\mathbf{E}]_{j,k} &= \left\{ \begin{array}{ll} 1, & c(j) = c(k) \ \mathrm{and} \ j \neq k \\ 0 , & c(j) \neq c(k) \ \ \mathrm{or} \ \ j = k \end{array} \right. .
\end{align}
\end{subequations}
A localized density matrix does not reduce the variance of this ensemble because $\mathbf{E}$ excludes its matrix elements.
An explicit local calculation thus has no effect and is unnecessary.

In the homogeneous spatial decay model, these two hybrid algorithms give similar error estimates.
Explicit hybrid $\mathbf{M}\odot\mathbf{P}$ and $\mathbf{M}\odot\mathbf{Q}$ approximations
 combine Eqs.\@ (\ref{local_density}) and (\ref{random_density}) into
\begin{equation} \label{hybrid_density}
 \mathbf{M}\odot\tilde{\mathbf{X}}_H \equiv \mathbf{M}\odot\tilde{\mathbf{X}}_L + \frac{1}{s} \sum_{i=1}^s  \mathbf{M}\odot [ (\mathbf{X} - \tilde{\mathbf{X}}_L) \mathbf{R}_i \mathbf{R}_i^\dag ] .
\end{equation}
We assume that local calculations follow the methodology in Sec.\@ \ref{local_approx},
 but previous hybrid calculations used chemically motivated saturation of bonds to embed molecular fragments for local calculations \cite{Baer_hybrid}.
For the random-phase ensemble, $r$ is replaced by $\max\{r,r_{\max}\}$ in the model density in Eq.\@ (\ref{random_homogeneous_bound}),
\begin{equation} \label{hybrid1_homogeneous_bound}
 \left\| \mathbf{M}\odot(\mathbf{P} - \tilde{\mathbf{P}}_{H1}) \right\|_F \sim \frac{\exp(-\gamma_{\min} r_{\max} )}{r_{\max}^{(D+1)/2}} \sqrt{ \frac{n r_0^D r_{\max}^D}{s r_{\min}^{2D} }} .
\end{equation}
Both $[r_{\min},r_{\max}]$ and $[r_{\max},\infty]$ intervals contribute roughly the same amount to the integral,
 thus better local approximations can only reduce this error estimate by a fixed fraction.
For the multi-color ensemble, $q \sim r_{\max}^D / r_{\min}^D$ and the $r_{\min}^{-D/2}$ prefactor in Eq.\@ (\ref{random_homogeneous_bound}) is kept,
 but the integral is rescaled from $r_{\min}$ to $r_{\max}$,
\begin{equation} \label{hybrid2_homogeneous_bound}
 \left\| \mathbf{M}\odot(\mathbf{P} - \tilde{\mathbf{P}}_{H2}) \right\|_F \sim \frac{\exp(-\gamma_{\min} r_{\max} )}{r_{\max}^{(D+1)/2}} \sqrt{ \frac{n r_0^D}{s r_{\min}^D} },
\end{equation}
 which is equal to the local error in Eq.\@ (\ref{local_homogeneous_bound}) for $s=1$.
Again, similar results hold for $\tilde{\mathbf{Q}}_{H}$ with an extra factor of $r_{\max}^{-1}$.
These two hybrid algorithms have identical error estimates when the number of random-phase samples is rescaled by $q$.

The cost analysis of hybrid calculations is complicated by the presence of two parameters, $s$ and $r_{\max}$,
 that distribute the computational burden over local and random parts.
To relate cost and error, we minimize an error per basis function $\varepsilon_{\mathrm{hybrid}}$ over $s$ and $r_{\max}$ at fixed cost.
The cost prefactor of multi-color ensembles is the total number of random vectors, $s q$,
 and the fixed-cost error for $T=0$, $\varepsilon_{\mathrm{hybrid}} \sim r_{\max}^{-1/2}$, is minimized when $s = 1$.
In this case, sampling-based variance reduction is an inefficient way to reduce errors.
If the hybrid random-phase ensemble costs are a linear combination of $r_{\max}^D$ and $s$ for the local and random calculations,
 then $s \sim r_{\max}^D$ is the fixed-cost error minimizer at $T=0$.
In both cases, the cost scaling does not improve over simpler local calculations.
At best, hybrid calculations can remove bias from local errors.
This analysis ignores the cost of directly evaluating $\tilde{\mathbf{X}}_L \mathbf{R}_i$ in Eq.\@ (\ref{hybrid_density}) for an
 explicit hybrid calculation, which adds an $O(s r_{\max}^D)$ prefactor but with a relatively small weight.
When this cost becomes relevant at high accuracy, the explicit hybrid algorithm is at a disadvantage.
Thus we use the multi-color vector ensemble to benchmark the effectiveness of hybrid algorithms.

The critical problem with existing hybrid algorithms is that the numerous, small off-diagonal matrix elements
 of $\mathbf{P}$ and $\mathbf{Q}$ contribute too much to the sampling variance of their random parts.
Even with perfect local approximations, $\tilde{\mathbf{X}}_L = \mathbf{M} \odot \mathbf{X}$,
 there can be a lot of residual variance in probing $\mathbf{X} - \tilde{\mathbf{X}}_L$ with random vectors.
Random approximations would complement local approximations better if their sampling variance
 was set by a localized sparsity pattern (e.g.\@ $\mathbf{E} = \mathbf{M}$).
This is possible, but it requires $O(n)$ columns in $\mathbf{R}_i$ and is incompatible with a linear-scaling cost unless other approximations are made.

\subsection{Analytical free-energy derivatives\label{analytic_derivative}}\vspace*{-0.3cm}

For zero-temperature electronic structure calculations, it is common to calculate observables from analytical derivatives of the total energy.
The finite-temperature analog is analytical derivatives of the free energy in Eq.\@ (\ref{free_energy}) for calculations of the observables in Eq.\@ (\ref{observable}).
The local and random approximations of Eq.\@ (\ref{observable}) that were considered in the previous subsection are
 not consistent with analytical derivatives of local and random approximations of Eq.\@ (\ref{free_energy}).
Because the free-energy matrix in Eq.\@ (\ref{decay_model}) has faster spatial decay than the density and energy-density matrices,
 it is more robust against localization errors and has less variance in the multi-color vector ensemble \cite{Barros_hybrid}.
While their implementation is more complicated, we expect analytical derivatives of approximate free energies
 to be more accurate than direct approximations of observables.

A simple but non-symmetric elementary operation for free-energy calculations is a decomposition of Eq.\@ (\ref{free_energy}) into
\begin{equation}
  F(\mathbf{y}) \equiv \mathbf{y}^\dag g(\mathbf{H} \mathbf{S}^{-1}) \mathbf{y} .
\end{equation}
In random calculations, we would average $F(\mathbf{y})$ over $\mathbf{y}$ drawn from a random vector ensemble.
In local calculations, $\mathbf{H}$ and $\mathbf{S}$ in $F(\mathbf{y})$ would be approximated with their local restrictions.
For methodological completeness, we review analytical free-energy derivatives
 of polynomial and rational approximations to $F(\mathbf{y})$.
However, we refrain from numerical implementation and testing in this paper because we presently lack
 an optimal rational approximation of $g(x)$ that is necessary for thorough and fair comparisons with a polynomial approximation.

The primitive of polynomial free-energy calculations is
\begin{equation}
 \tilde{F}_{\mathrm{poly}}(\mathbf{y}) \equiv \sum_{j=0}^{p-1} \beta_j \mathbf{y}^\dag C_j(\mathbf{H} \mathbf{S}^{-1}) \mathbf{y} \approx F(\mathbf{y})
\end{equation}
 for Chebyshev coefficients $\beta_k$.
Its analytical derivative is
\begin{subequations}
\begin{align} \label{polynomial_response}
 \frac{d \tilde{F}_{\mathrm{poly}}}{d \lambda} (\mathbf{y}) &= \sum_{j=0}^{p-2} \mathbf{y}^\dag C_j(\mathbf{H} \mathbf{S}^{-1})  \left[ \frac{d \mathbf{H}}{d \lambda} - \mathbf{H}  \mathbf{S}^{-1} \frac{d \mathbf{S}}{d \lambda} \right] \mathbf{S}^{-1} \mathbf{x}_j , \\
 \mathbf{x}_j &= (2 - \delta_{j,0}) \sum_{k=j+1}^{p-1} \beta_k U_{k-j-1}(\mathbf{H}\mathbf{S}^{-1}) \mathbf{y},
\end{align}
\end{subequations}
 which is derived from generating functions of the Chebyshev polynomials
 including Chebyshev polynomials of the second kind, $U_{k-1}(\cos \theta) \equiv \sin( k\theta ) / \sin (\theta)$, to simplify the result.
It is similar to the outcome of automatic differentiation \cite{Barros_hybrid}, and 
 both require a careful selection of intermediates for efficient evaluation.
Here, it is efficient to precompute $\mathbf{x}_0$ and compute subsequent $\mathbf{x}_j$ by subtracting one $U_{j-1}(\mathbf{H}\mathbf{S}^{-1}) \mathbf{y}$ at a time.

The primitive of rational free-energy calculations is
\begin{equation}
 \tilde{F}_{\mathrm{rational}}(\mathbf{y}) \equiv 2 \, \mathrm{Re} \sum_{j=1}^p \tau_j \mathbf{y}^\dag \mathbf{S} \mathbf{G}(\omega_j) \mathbf{y} \approx F(\mathbf{y})
\end{equation}
 for residues $\tau_k$ and poles $\omega_k$.
Its analytical derivative is
\begin{align}\label{rational_derivative}
 \frac{d \tilde{F}_{\mathrm{rational}}}{d\lambda}(\mathbf{y}) &= 2 \, \mathrm{Re} \sum_{j=1}^p \tau_j \mathbf{y}^\dag \mathbf{S} \mathbf{G}(\omega_j) \left[ \omega_j \frac{d \mathbf{S}}{d\lambda} - \frac{d \mathbf{H}}{d\lambda} \right] \mathbf{G}(\omega_j) \mathbf{y} \notag \\
 & \ \ \ \ +  2 \, \mathrm{Re} \sum_{j=1}^p \tau_j \mathbf{y}^\dag \frac{d \mathbf{S}}{d\lambda} \mathbf{G}(\omega_j) \mathbf{y} .
\end{align}
Each pole in the rational approximation contributes a separate term to the derivative,
 while the polynomial approximation in Eq.\@ (\ref{polynomial_response}) has cross terms between different basis functions.

We can estimate errors in analytical free-energy derivatives using the homogeneous spatial decay model.
For an accurate rational approximation of $g(x)$ and local models of $\mathbf{G}(\omega_i)$ that are analogous to Eq.\@ (\ref{homogeneous_local_density}),
 $\mathbf{P}$ as calculated from Eq.\@ (\ref{rational_derivative}) has a leading-order local approximation error of
\begin{align}
 [\mathbf{P} - \tilde{\mathbf{P}}_{L'}]_{j,k} &\sim \sum_{i=1}^p \tau_i [ \mathbf{G}(\omega_i) \boldsymbol( \mathbf{G}(\omega_i) - \tilde{\mathbf{G}}_L(\omega_i) \boldsymbol) ]_{j,k} \notag \\
 &\sim \sum_{l=1}^n \frac{\exp[-\gamma_{\min} (|\mathbf{r}_j - \mathbf{r}_l | + \max\{|\mathbf{r}_l - \mathbf{r}_k |,r_{\max}\})] }{(|\mathbf{r}_j - \mathbf{r}_l | + \max\{|\mathbf{r}_l - \mathbf{r}_k |,r_{\max}\})^2} \notag \\
 & \ \ \ \ \ \ \times \frac{1}{(|\mathbf{r}_j - \mathbf{r}_l | \max\{|\mathbf{r}_l - \mathbf{r}_k |,r_{\max}\})^{(D-1)/2} } ,
\end{align}
 assuming that the sum over poles reconstructs the asymptotic behavior of the homogeneous model of $g(\mathbf{H}\mathbf{S}^{-1})$ in Eq.\@ (\ref{decay_model}).
If we also assume that all near-diagonal matrix elements have a similar magnitude, then the overall error is
\begin{align} \label{derivative_homogeneous_bound}
 \left\| \mathbf{M}\odot(\mathbf{P} - \tilde{\mathbf{P}}_{L'}) \right\|_F &\sim \sqrt{ \frac{n r_0^D}{r_{\min}^D} \int_0^{\infty} \frac{\exp[-2\gamma_{\min} (r+r_*)] }{(r+r_*)^4 (r r_*)^{D-1}} \frac{r^{D-1} dr}{r_{\min}^D} } \notag \\
  &\sim \frac{\exp(-2 \gamma_{\min} r_{\max} )}{r_{\max}^{(D+2)/2}} \sqrt{ \frac{n r_0^D}{ r_{\min}^D} }
\end{align}
 for $r_* \equiv \max\{ r, r_{\max} \}$.
It is smaller than local error estimates in Eq.\@ (\ref{local_homogeneous_bound}) 
 and similar to random error estimates for analytical derivatives in the multi-color vector ensemble \cite{Barros_hybrid}.

\subsection{Implementation details}\vspace*{-0.3cm}

All of the relevant theoretical details have been articulated,
 and we now discuss some implementation details that may be of technical interest.
This discussion is not complete enough to characterize the software fully,
 and we provide open access to the source code \cite{source_code}
 for readers who are interested in details beyond the limited scope of this subsection.

The software implementation is limited to shared-memory parallelism and is intended for use on a workstation or single supercomputer node.
It is written in C and uses OpenMP for thread-based parallelism everywhere except the PEXSI-based solver that uses MPI for compatibility with PEXSI.
External dependencies of the software are limited to standard libraries (BLAS, LAPACK, and FFTW3) and PEXSI.
All benchmarks are run in the same computing environment, which is limited to 16 cores, 64 GB of memory, and 48 hours of wall time.
For simplicity, we assign one OpenMP thread or MPI task to each core.
The wall-time restriction is not a fundamental hardware limitation, but it serves in Fig.\@ \ref{scaling_fig} as an example of when finite computational resources
 prevent large-prefactor, low-scaling algorithms from reaching their asymptotic scaling regime.

All pre- and post-processing steps are implemented with a linear-scaling cost, emphasizing simplicity over performance.
The simulation volume is partitioned into uniform cubes and atoms are assigned to these cubes for efficient construction of neighbor lists.
This is efficient when the spatial distribution of atoms is approximately uniform.
To assign colors to atoms for a random-vector ensemble,
 we apply the Welsh-Powell vertex coloring algorithm \cite{Welsh_Powell} to a graph with atoms as vertices
 and interatomic distances below a target threshold as edges.
Each atomic orbital is also assigned to its own color, therefore each atom color corresponds to nine basis-function colors.

All matrices are stored in a block-sparse matrix format and all vectors are stored in a compatible block format to improve performance of matrix and vector operations.
For simplicity, we constrain blocking to the 9-by-9 matrix blocks associated with each atomic basis set rather than optimize block size for performance.
This results in the performance of matrix-vector multiplication being limited by memory bandwidth instead of processor speed.
All vectors are grouped into sets of nine so that matrix-vector multiplication is decomposed into 9-by-9 block matrix operations using BLAS.
While we solve linear systems with nine right-hand-side vectors simultaneously, we use standard iterative solvers and not block solvers.
Matrices are converted into a dense format when passed into LAPACK and a nonblocked sparse format when passed into PEXSI.

To study the limit of infinite cluster size, we implement two algorithms for periodic boundary conditions.
A conventional cubic-scaling algorithm utilizes Bloch's theorem and uniform sampling of the Brillouin zone,
 and a localized linear-scaling algorithm calculates columns of the density matrix associated with the central periodic unit cell without any use of Bloch's theorem.
This functionality is primarily used for convergence studies of parameters such as the localization radius.

The only observables using Eq.\@ (\ref{observable}) that we implement are atomic forces and the stress tensor for periodic systems.
This is facilitated by the simple analytical form of matrix elements in the NRL tight-binding model.

\section{Results\label{results}}\vspace*{-0.3cm}

We now compare the computational costs of four electronic structure algorithms 
 with linear (localized and randomized), quadratic (PEXSI), and cubic (LAPACK) system-size scaling
 on copper clusters as a function of their size and temperature.
We quantify implementation-specific performance overheads that limit the accuracy of these comparisons,
 optimize several algorithm parameters to attain accuracy targets for calculated observables,
 and fit cost models to benchmark data.

\begin{figure}[t!]
\includegraphics{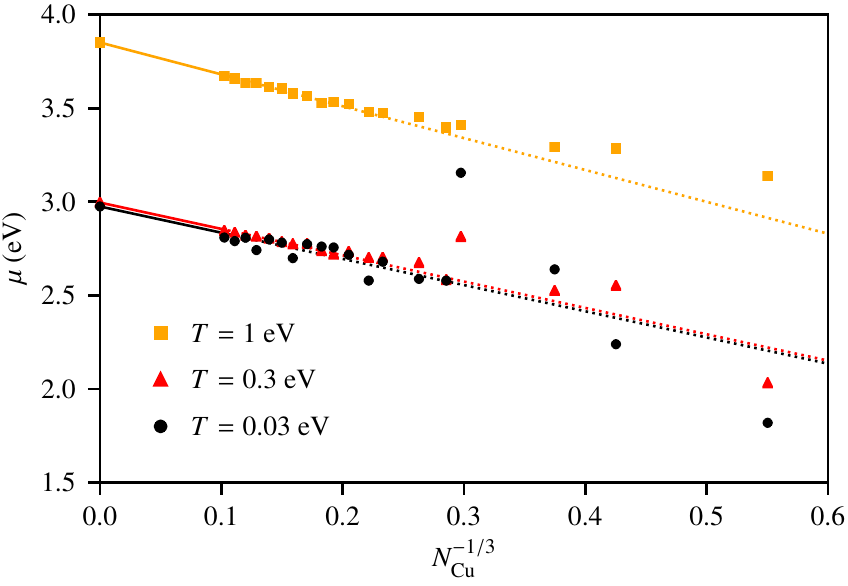}
\caption{\label{chemical_potential_fig} Optimized chemical potentials of simulated copper clusters with $N_{\mathrm{Cu}}$ atoms at a temperature $T$
 and the chemical potential model in Eq.\@ (\ref{mu_model}) used for $N_{\mathrm{Cu}} \ge 1,000$ (solid and dotted lines). }
\end{figure}

We perform all benchmarking on a simple structural model.
Copper clusters are extracted from the bulk fcc crystal with a lattice constant of 3.52 $\AA$
 by retaining atoms inside a masking polyhedron.
We use a rhombicuboctahedron to expose \{100\}, \{110\}, and \{111\} surfaces of the crystal.
Clusters are centered on either an atom or an octahedral interstitial site.
We do not relax these cluster geometries.
For clusters smaller than 1,000 atoms,
 we use optimized chemical potentials shown in Fig. \ref{chemical_potential_fig}
 and approximate the chemical potential of larger clusters as
\begin{align}
 \mu(N_{\mathrm{Cu}},T) & \approx \left[ 2.975 - 1.4 N_{\mathrm{Cu}}^{-1/3} \right] \mathrm{eV} \notag \\
 & \ \ \ \ + \exp(-1.6 \mathrm{\, eV} / T) \left[ 4.34 - 1.5 N_{\mathrm{Cu}}^{-1/3} \right] \mathrm{eV} \label{mu_model}
\end{align}
for $N_{\mathrm{Cu}}$ atoms and temperature $T$.
These benchmarks act as a representative of a single iteration in a self-consistency cycle
 that optimizes geometries and chemical potentials.
The outer parameter optimization process introduces complications and
 computational challenges that we do not address here.

\subsection{Performance overheads\label{performance}}\vspace*{-0.3cm}

We want benchmarks to be accurate representations of the relative costs of multiple algorithms
 that minimize the impact of implementation-specific details.
This is straightforward for the algorithms studied in this paper because their costs are all dominated
 by a single computational bottleneck with a mature implementation.
The main concern of multi-core benchmarks is the strong scaling of costs with the number of cores,
 which biases towards implementations with better strong scaling.
It becomes increasingly difficult to maintain good strong scaling as more cores become available,
 and our 16-core limit avoids the regime where memory bandwidth becomes a problem.

\begin{figure}[t!]
\includegraphics{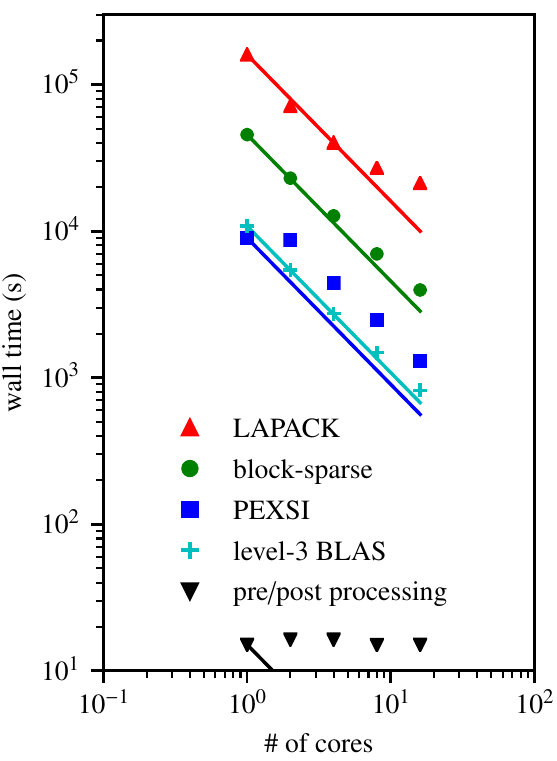}
\caption{\label{strong_scaling_fig} Strong scaling of three computational bottlenecks compared
 with pre/post-processing costs and a reference level-3 BLAS dense matrix-matrix multiplication of equivalent size (52,137).
Guide lines show ideal strong scaling behavior relative to the serial performance.}
\end{figure}

We visualize a strong-scaling test in Fig. \ref{strong_scaling_fig} for an arbitrarily
 chosen 5,793-atom Cu cluster at $T = 1$ eV with a $3$-pole-pair approximation of $f(x)$ for PEXSI and localized calculations.
We compare total simulation times as the pre/post-processing costs of the NRL tight-binding model are negligible
 for small numbers of cores, even without efficient threading.
All of the algorithm implementations show reasonable strong scaling to 16 cores with only minor performance degradations.

For the LAPACK-based algorithm, the 50\% performance degradation at 16 cores
 results from its reliance on a reduction to tridiagonal form for matrix diagonalization.
This reduction uses level-2 (matrix-vector) BLAS operations that have more difficulty in hiding memory bottlenecks under arithmetic costs
 compared to level-3 (matrix-matrix) BLAS operations.
While there has been progress in fixing this problem for distributed-memory parallelism \cite{ELPA},
 it has not yet been fully adapted and implemented for strictly shared-memory parallelism.

For the PEXSI-based algorithm, we observe strong scaling
 that is good but offset from the serial performance.
This may be the result of a communication overhead of the distributed-memory parallel algorithm
 in the absense of a shared-memory implementation.
The interface to PEXSI does not use a block-sparse matrices,
 but it internally blocks sparse matrices during factorizations performed by the SuperLU library \cite{PEXSI_details}.
Thus it should not have much of a performance overhead relative to
 a block-sparse interface beyond matrix format conversions.

For the block-sparse-matrix-based algorithms, the physical 9-by-9 block size of an $spd$ atomic orbital basis set
 maintains good strong scaling up to 16 cores.
The implementations are threaded over small, unthreaded block-matrix operations.
For larger numbers of cores, a larger block size would be needed.
Changing block sizes is a nontrivial partitioning problem that 
 wastes memory and computation on nonzero matrix elements
 to distribute work more efficiently over many threads.

Successful strong scaling does not necessarily indicate that computational resources have been efficiently utilized.
Large dense matrix-matrix multiplications operate at 170 Gflops on 16 cores,
 while our block-sparse matrix-based solvers operate at only 20 Gflops.
This discrepancy in performance highlights the difficulty of comparing benchmarks of new algorithms to established algorithms with mature implementations.
There is still value in performing such comparisons, but they should be presented with appropriate caveats.
In this case, the caveat is the non-optimal block size in block-sparse matrix operations,
 which may be alleviated by local clustering of atoms.

\subsection{Algorithm optimizations\label{preconditioners}}\vspace*{-0.3cm}

\begin{figure}[t!]
\includegraphics{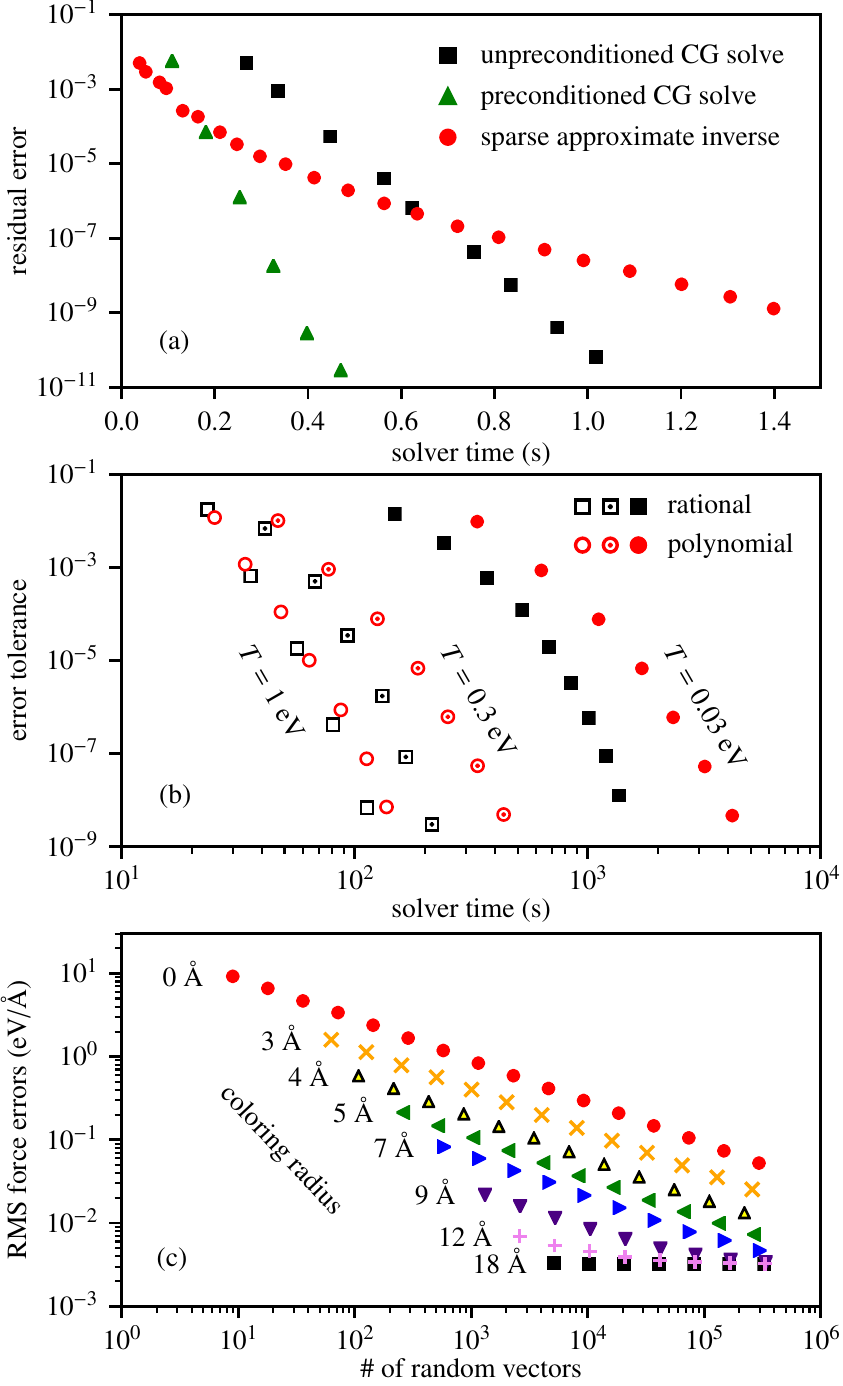}
\caption{\label{optimization_fig} Algorithm optimizations are informed by performance tests demonstrating that
 (a) preconditioning of iterative linear solvers only reduces solving times by half in the best-case scenario of the overlap matrix,
 (b) with no preconditioner, rational approximations are more efficient than polynomial approximations for $f(\mathbf{H}\mathbf{S}^{-1}) \mathbf{x}$ calculations, and
 (c) increasing the coloring radius that defines multi-color vector ensembles is more effective than adding more random samples.
}
\end{figure}

To simplify the benchmarks, we do not attempt to test each algorithm variant discussed in Sec.\@ \ref{methods}.
Instead, we make three design decisions based on some representative tests that focus
 our attention on the best-performing algorithms.
These three tests and the decisions that they guide are shown in Fig. \ref{optimization_fig}.

For localized electronic structure calculations, the natural choice of preconditioner is a sparse approximate inverse \cite{SAI}.
Localized calculations of $\mathbf{S}^{-1} \mathbf{e}_i$ and $(\mathbf{H} - \omega \mathbf{S})^{-1} \mathbf{e}_i$
 can be used to construct these preconditioners one column at a time.
We further restrict their sparsity to balance cost and accuracy and symmetrize them for convenience.
Fig.\@ \ref{optimization_fig}a shows the tradeoff between cost and accuracy
 and the benefit of preconditioning an iterative evaluation of $\mathbf{S}^{-1} \mathbf{x}$
 for a periodic calculation with a localization radius of 30 \AA.
Every circle represents a sparse approximate inverse with a restricted localization radius from 7 \AA \ and 28 \AA \ in increments of 1 \AA.
We truncate at 7 \AA \ for the preconditioner, which restricts it to the sparsity pattern of $\mathbf{S}$.
At low accuracy, direct use of a sparse approximate inverse is the fastest way to evaluate $\mathbf{S}^{-1} \mathbf{x}$
 because it avoids calculating the residual error as in an iterative solver.
Preconditioning is successful in this case, reducing solver times by half at a cost of increased memory usage in storing the preconditioner.
The small condition number of $\mathbf{S}$, $\kappa \approx 5$, means that there is little
 opportunity for a preconditioner to improve performance.

Unfortunately, sparse approximate inverse preconditioners degrade in performance for $\mathbf{H} - \omega \mathbf{S}$ at low temperature
 as the imaginary part of $\omega$ becomes small.
Our initial strategy was to order the linear solves by decreasing $|\mathrm{Im}(\omega)|$ and construct a sparse approximate inverse from each $\omega$
 to precondition the next $\omega$ value.
For too much truncation, these preconditioners fail to reduce condition numbers and behave erratically.
This strategy works at $T=0.3$ eV, where the preconditioner still can be truncated reliably with the sparsity pattern of $\mathbf{H} - \omega \mathbf{S}$,
 but fails at $T = 0.03$ eV because the preconditioner requires a large localization radius to reduce the condition number.
A more modest strategy is to use a single preconditioner defined by $\omega_{\mathrm{pre}} = \mu + i \pi T_{\mathrm{pre}}$ for every $\omega$,
 which enables more control over sparsity, but limits the worst-case reduction in condition number from Eq.\@ (\ref{condition_number0}) to $\kappa \approx T_{\mathrm{pre}}/T$.
In principle, this fails to alter the $T$-dependence of condition numbers.
In practice, we can at best reduce the solver time in half by fine-tuning $T_{\mathrm{pre}}$ and the preconditioner localization radius for each value of $T$.
Ultimately, we decide against using preconditioners because the modest reduction in solver times does not justify
 the extra fine-tuning and memory usage that is required.
The largest of our benchmarks are memory limited, thus memory reduction
 is our optimization priority when testing large systems.

Choosing between polynomial and rational approximations of the Fermi-Dirac function $f(x)$ is more straightforward.
We compare their costs of approximating $f(\mathbf{H}\mathbf{S}^{-1}) \mathbf{x}$ in Fig.\@ \ref{optimization_fig}b on a 5,793-atom cluster
 as a function of the error tolerance in the iterative-solver residuals and function approximations.
In this application, rational approximations have a clear performance advantage that grows with decreasing $T$ and errors.
Rational approximations require a small number of poles,
 and most of the computational effort is spent on the pole closest to the real axis whereby $\mathrm{Im}(z_i) \approx \pi T$.
Unpreconditioned iterative solvers effectively approximate $(\mathbf{H} - \omega \mathbf{S})^{-1}$ as polynomials in $\mathbf{H} - \omega \mathbf{S}$,
 but the approximation is adapted to the details of its spectrum rather than a uniform approximation over $[\epsilon_{\min}-\mu, \epsilon_{\max}-\mu]$.
Also, rational approximations can benefit more from efficient preconditioning.
Sparse approximate inverse preconditioners benefit polynomial and rational approximations similarly,
 and this assessment is the same whether they are used or not.

The choice of random vector ensemble for the randomized algorithms is also straightforward.
For a 1,192-atom cluster at $T=1$ eV, we vary the coloring radius and number of samples of a multi-color vector ensemble in Fig.\@ \ref{optimization_fig}c.
The leftmost point for each symbol corresponds to one sample.
Consistent with the analysis in Sec.\@ \ref{hybrid_approx}, it is more effective to increase the coloring radius
 and number of vectors per sample rather than increase the number of samples to reduce the finite-sampling errors.
The error floor occurs when finite-sampling errors are driven below other errors in the calculation that cause bias in the overall error.
We run all benchmarks in the single-sample limit and tune the coloring radius to adjust errors.

\subsection{Error calibration\label{self_averaging}}\vspace*{-0.3cm}

Algorithms that reduce computational costs by introducing multiple approximations
 often have multiple error tolerances that must be tuned to balance cost and accuracy.
Our linear-scaling algorithms control errors through four parameters for
 function approximation errors, iterative-solver residual errors,
 localization radius, and coloring radius.
Our goal is to achieve typical levels of convergence in standard observables of 0.01 eV/atom for total energy, 0.01 eV/\AA \ for forces, and 1 GPa for stresses.
We also test the correlation between these errors and density-matrix errors as in Eq.\@ (\ref{error_bound})
 to consolidate them into a single density-matrix error target.
We assume that parameter-dependent errors are uncorrelated and tune them individually while holding other parameters at over-converged values.

\begin{figure}[t!]
\includegraphics{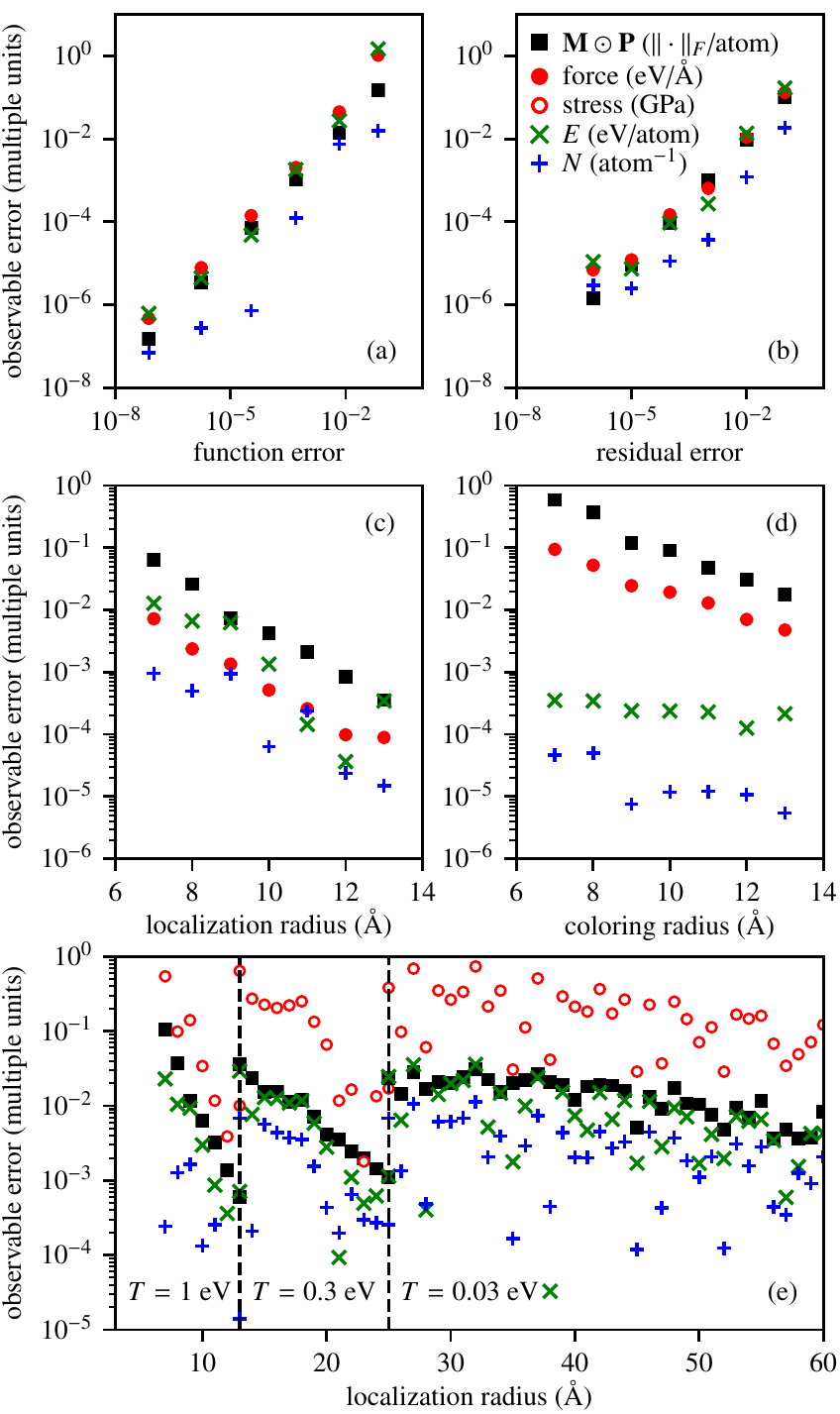}
\caption{\label{error_fig} Observable errors generated by (a) function approximation, (b) iterative-solver residuals, (c) localization, and (d) finite sampling.
We calibrate the localization radius at each simulated temperature by using periodic calculations of bulk copper (e) as a reference.
}
\end{figure}

The function approximation errors in Fig.\@ \ref{error_fig}a and iterative-solver residual errors in Fig.\@ \ref{error_fig}b
 are well-controlled numerical errors that are directly specified by tolerances and predictably propagate into observable errors.
They contribute to the cost prefactor as $O(\log 1/\epsilon)$ for a tolerance $\epsilon$, which allows for an efficient convergence to high accuracy if required.
These tests were performed on a 116-atom cluster at $T = 0.3$ eV.

The localization radius in Fig.\@ \ref{error_fig}c and the coloring radius in Fig.\@ \ref{error_fig}d are parameters that indirectly control errors.
We have rationalized this behavior with a model of simple metals that applies to Cu clusters,
 but the general behavior will be more complicated and convergence is expensive for low $T$.
These tests were performed on a 5,793-atom cluster at $T = 1$ eV to observe a clear exponential decay of error with radius.
While the analysis in Sec.\@ \ref{trace_approx} finds that localized and randomized algorithms should have the same error decay,
 we observe that the error prefactor is ten times larger in the randomized case.
We do not have a simple explanation for this observation, but we find that it is relatively insensitive to $T$.

We use localized periodic calculations in Fig.\@ \ref{error_fig}e to tune the localization and coloring radii
 even when they are too large to compute in non-periodic benchmarks.
The observed trend is consistent with Fig.\@ \ref{NRL_fig}d, where convergence becomes more erratic as the density matrix transitions from an exponential to algebraic decay at low $T$.
Without exponential decay, our ability to keep both costs and errors under control is severely degraded.
There are frameworks \cite{NTPoly} for approximate sparse matrix algebra that impose sparsity by truncating small matrix elements in intermediate operations,
 but this tighter control of errors will inevitably cause an uncontrollable loss of sparsity as temperature decreases.
It is no longer practically useful to treat the density matrix as a sparse matrix in this regime.

The apparent correlation between density matrix errors and other observable errors is relatively consistent in Fig.\@ \ref{error_fig}
 with the exceptions of $E$ and $N$ in Fig.\@ \ref{error_fig}d.
These are examples of self-averaging errors in randomized calculations.
Their errors behave as in Eqs.\@ (\ref{sampling_error}) and (\ref{estimate_bound}).
As previously observed \cite{random_comment}, the Frobenius norms that determine sampling variance have a different dependence on system size
 for system-averaged and local intensive observables.
The matrix of a system-averaged observable has $O(n)$ eigenvalues of size $O(1/n)$ for $n$ atoms,
 which results in an $O(1/n)$ variance.
In contrast, the matrix of a local observable has $O(1)$ eigenvalues of size $O(1)$, which results in an $O(1)$ variance.
The same scaling behavior occurs in the bound for deterministic errors in Eq.\@ (\ref{error_bound}),
 but it is too loose of a bound to enforce this behavior in practice.

\subsection{Scaling benchmarks}\vspace*{-0.3cm}

We benchmark the localized and randomized linear-scaling electronic structure algorithms
 to identify the empirical onset of their asymptotic linear-scaling costs
 and crossover points with the more established quadratic-scaling and cubic-scaling algorithms.
The test set is Cu clusters that are logarithmically distributed from 6 to 147,570 atoms
 with the ratio of atoms in consecutive clusters of $\approx$$2^{1/3}$ so that the diagonalization cost is approximately doubled with every successive cluster.
Over the target set of temperatures in eV, $\{ 1, 0.3, 0.03 \}$,
 we set the number of pole-pairs in the rational approximation to $\{ 3, 4, 6 \}$
 to reduce the approximation error to $10^{-3}$ and similarly set the residual error tolerance to $10^{-3}$
 as guided by the error analysis in the previous subsection.
The choice of radii has the largest effect on cost, therefore we choose the minimal values
 that we expect to be required for convergence, $\{8,14,38\}$ in \AA \ for the localization radii and  $\{12,22,96\}$ for the coloring radii.

\begin{figure*}[p!]
\includegraphics{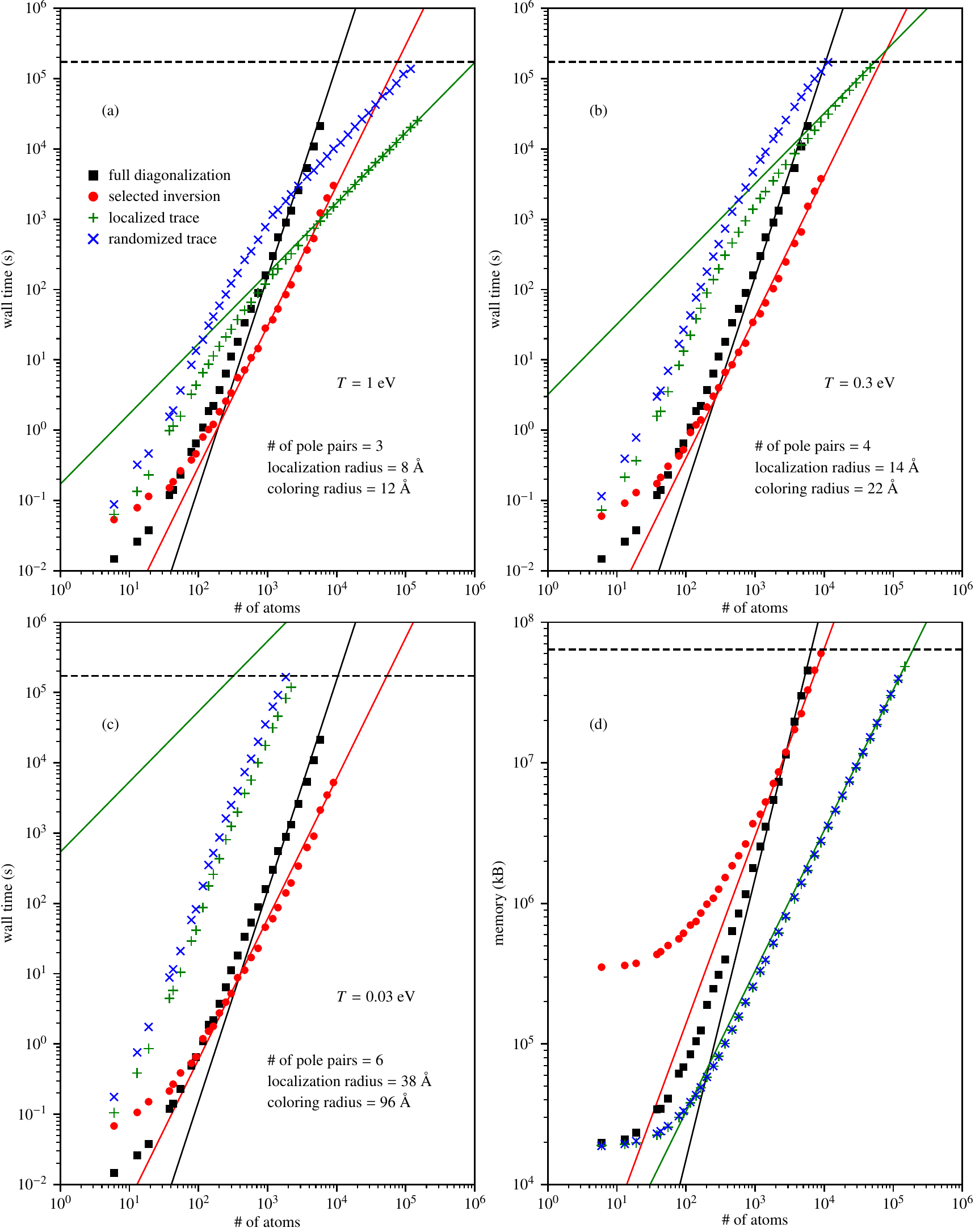}
\caption{\label{scaling_fig} Performance benchmarks of rhombicuboctahedral copper clusters at (a) $T=1$ eV, (b) $T=0.3$ eV, and (c) $T=0.03$ eV
 and (d) their $T$-independent memory usage on a 16-core computer limited to 64 GB of memory and a 48-hour wall time (dashed lines).
 Guide lines show the theoretical asymptotic scaling of costs with prefactors fit to data (the linear-scaling asymptote at $T=0.03$ eV is fit to periodic calculations).
}
\end{figure*}

The costs observed in Fig. \ref{scaling_fig} agree well with expectations.
Matrix diagonalization has an $O(n^3)$ time and $O(n^2)$ memory cost for $n$ atoms
 that is independent of temperature and matrix structure.
Selected inversion has an $O(\log(1/T) n^{3-\min\{3/D,2\}})$ time and $O(n^{2-\min\{2/D,1\}})$ memory cost
 at temperature $T$ and in $D$ spatial dimensions,
 with additional cost prefactors that are dependent on details about the $\mathbf{H} - \omega \mathbf{S}$ sparsity pattern.
The quadratic-cubic crossover point below 100 atoms is caused by
 the simplicity of minimal-basis tight binding with aggressive matrix element localization,
 and it will increase substantially for more complicated models.
Both linear-scaling algorithms have an $O(r_{\max}^D n/T)$ time and $O(n)$ memory cost where $r_{\max}$
 is the localization or coloring radius.
The increase of both $r_{\max}$ and the condition number contribute to the large time increase with decreasing $T$.
The linear-quadratic crossover is sensitive to $T$, and it is many 1,000's of atoms in the best-case scenario of $T=1$ eV
 and estimated at $10^7$ atoms for $T = 0.03$ eV.

There are several anomalies about memory usage in Fig.\@ \ref{scaling_fig} that are worth noting.
First, the implementations in this paper have no $T$ dependence on memory usage
 because poles of the rational approximation are computed in serial.
Parallelization over poles could introduce a small $T$ dependence.
The simple linear-scaling algorithms studied in this paper have a minimal memory footprint
 that is dominated by storage of the essential sparse-matrix inputs and outputs.
More efficient linear-scaling algorithms are likely to need more memory to store structured approximations of $(\mathbf{H} - \omega \mathbf{S})^{-1}$
 as preconditioners or for faster algorithms that approximately invert sparse matrices.
Finally, the memory usage of small clusters is dominated by computer overhead,
 particularly for MPI parallelization in PEXSI.

We also note that the costs of the linear-scaling algorithms are highly predictable after a limited amount of preprocessing
 to determine the average number of nonzero matrix elements per atom.
The cost per block-sparse matrix-vector product can be estimated from benchmarking individual block operations,
 and the number of products can be estimated from bounds on the condition number based on estimates of $\epsilon_{\min}$ and $\epsilon_{\max}$.
This is important for deciding on which is the fastest algorithm for a specific problem instance.
While we do not thoroughly test such cost predictions, we check that localized calculations of bulk copper are
 an accurate predictor of the cost per atom at $T=0.3$ eV and use them to predict
 the unattained asymptotic localized linear-scaling cost at $T=0.03$ eV in Fig.\@ \ref{scaling_fig}c.

\section{Discussion\label{discussion}}\vspace*{-0.3cm}

Within a limited set of tests using one model system and a few algorithms,
 we can make several new observations about performance.
Selected inversion (PEXSI) has a robust regime over several orders of magnitude in system size
 where it beats cubic-scaling diagonalization and the available linear-scaling algorithms even in three spatial dimensions.
This success of PEXSI further increases the crossover point for linear-scaling algorithms and limits their near-term applicability.
Memory becomes the bottleneck suprisingly quickly for all algorithms without distributed-memory implementations.
Linear-scaling algorithms achieve a serial performance target of $\sim$1 s/atom
 that is comparable to similar benchmarks listed in Sec.\@ \ref{intro}, but only at $T \ge 1$ eV for metallic systems.
While the benchmarks presented here rely on custom solver implementations,
 future benchmarks may utilize the electronic structure infrastructure (ELSI) project \cite{ELSI}
 for convenient access to a common set of mature solver implementations with minimal effort.

Although randomized electronic structure algorithms were first proposed 25 years ago \cite{original_random}
 and the subject of increased interest over the last 5 years \cite{random_quantum},
 a quantitative performance comparison with localized algorithms has not been published before this paper.
By exploiting localization through a multi-color vector ensemble \cite{Barros_hybrid},
 randomized algorithms match the performance of localized algorithms in Fig.\@ \ref{scaling_fig} within a factor of ten.
These calculations use only one random multi-vector sample because localization-based variance reduction
 is more effective than sampling-based variance reduction.
In contrast, we observe poor performance of the random-phase ensemble in Sec.\@ \ref{preconditioners}.
While it performs equally well for metallic and insulating systems, this performance is uniformly poor.
There are proposals \cite{Baer_hybrid} to operate randomized algorithms in a fast, large-error regime
 and to mitigate unbiased force errors as the random forces driving Langevin dynamics.
However, such a scheme does not mitigate electron-density errors, which will
 propagate into self-consistent potential errors, and the limited characterization of force errors
 is not sufficient to satisfy the fluctuation-dissipation relations exactly.
Randomization will likely be a useful tool in future algorithm development, but it
 should be compared and combined with other possible tools while carefully assessing the details of cost and accuracy.

It is typical to be conservative and overconverge numerical errors in published results of electronic structure simulations.
However, the high error-sensitivity of linear-scaling algorithm costs and extensivity of certain errors may require
 changes to standard practice and more consideration of errors.
If we want to calculate the total energy difference, $E(\lambda_2) - E(\lambda_1)$, over a reaction coordinate $\lambda$,
 we cannot rely on explicit calculations of $E(\lambda_1)$ and $E(\lambda_2)$ if a system is large
 and the errors in $E(\lambda)$ are extensive and uncorrelated.
A viable alternative might be to calculate $E'(\lambda)$ and integrate it from $\lambda_1$ and $\lambda_2$.
It also may be difficult to conserve energy over long molecular dynamics trajectories.
Accurate energy conservation might be possible if forces are taken from the analytical free-energy derivatives 
 discussed in Sec.\@ \ref{analytic_derivative} and a common pesudorandom sequence is used by randomized algorithms at every time step.
Another solution \cite{energy_thermostat} is to bias dynamics towards a target energy just as
 thermostats are used to maintain a constant temperature in molecular dynamics simulations.
Ultimately, it is wasteful for numerical errors to be many orders of magnitude smaller than the model errors (e.g.\@ density functionals) in a simulation.

Large-basis electronic structure calculations are nowadays more common
 than the minimal-basis calculations performed in this paper.
A good way to compare costs in these two cases is to introduce a basis set efficiency parameter $\alpha$
 that defines the number of basis functions per atom.
It can vary  from $\sim$10 for a minimal basis to $\sim$1,000 for a large basis.
Costs scale as $O(\alpha^2)$ for the localized linear-scaling algorithm
 and also for the randomized algorithm with a multi-color vector ensemble.
While some linear-scaling calculations are performed directly in a large basis \cite{large_local},
 it is more common to project the problem into a small basis of localized orbitals that span the occupied electronic states \cite{ONETEP_metal}.
The cost of this projection is $O(\alpha)$, and it can be further hidden if it is less expensive than solving the projected problem.
Conversely, projected calculations will not benefit from increasing efficiency of linear-scaling algorithms on the projected problem
 once their cost has become less than the projection cost.
Thus the priorities of method development are different for small-basis and large-basis applications.

Warm dense matter spans a temperature range from $\sim$1 eV to $\sim$100 eV
 and a characteristic free-electron decay length of $\sim$0.1 \AA \ to $\sim$1 \AA \ for density matrices \cite{warm_dense_matter}.
In this regime, the density matrix decays on length scales comparable to features in potential energy functions.
We expect that localized linear-scaling algorithms will have a very small crossover point here with conventional algorithms
 that use iterative eigensolvers in a large basis.
The number of partially occupied states that are conventionally calculated grows as $\propto T^{3/2}$ for temperature $T$,
 causing simulation times to increase as $O(T^{9/2})$.
The cost of localized linear-scaling algorithms will decrease as $O(T^{-3/2})$ according to the analysis in this paper,
 but it should asymptote to a $T$-independent cost when the length scale of density decay becomes smaller than features in the potential.
In this limit, it may be beneficial to develop efficient singularity models and high-$T$ expansions for electronic Green's functions.
Orbital-free DFT also enables efficient warm dense matter simulation \cite{OFDFT},
 but it needs approximate kinetic-energy functionals that are difficult to improve in accuracy.
Localized linear-scaling algorithms are a systematically improvable alternative.

The NRL tight-binding model used in this paper is a good compromise between simplicity and realism.
However, future linear-scaling algorithm development may benefit from even simpler models
 to reduce memory use, simplify bookkeeping and partitioning, and narrow focus to the essential challenges.
We propose that large clusters extracted from a simple-cubic lattice
 of an orthogonal tight-binding model with only nearest-neighbor hopping are sufficient tests if error targets are well calibrated.
We have seen that density-matrix errors are a good proxy for other observable errors.
An error target of between $0.01$ and $0.001$ in Frobenius norm per column for off-diagonal matrix elements
 within four to five lattice hops is comparable to the target used in this paper.
Since finite-$T$ effects are small at ambient conditions, we can target $T=0$ while adjusting $T$ to study the $T$-dependence of costs.
We can also vary $\mu$ and electron density, but they should not have much effect on cost.
The very sparse matrices of this model might artificially favor linear-scaling algorithms,
 but they first need to be effective in a controlled setting to be viable for more realistic models.

In the remainder of this section, we discuss two theoretical concepts that might be useful for
 further improving localized and randomized linear-scaling electronic structure algorithms
 and enhancing their compatibility in hybridized local-random algorithms.
We use the simple-cubic orthogonal tight-binding model as a numerical example to test both concepts.

\subsection{Localization self-energy}

The localized linear-scaling algorithm studied in this paper does not use global information in performing calculations on a local subsystem.
This fails to utilize the popular concept of embedding, whereby the global environment is approximated or modeled at a lower level of theory and not simply ignored.
For example, the environment can be modeled with a classical interatomic potential \cite{QMMM},
 dangling bonds of a subsystem can be saturated \cite{Baer_hybrid},
 representations of environmental effects as a continued fraction can be approximated \cite{recursion},
 local scattering methods model the environment as a uniform electron gas \cite{local_scattering},
 and the environment can be incorporated into a local Green's function through an embedding self-energy \cite{PEXSI_Sigma}.
These latter three examples are all based on Green's functions, which are particularly useful for embedding and our focus here.

The local subsystem calculations each construct one sparse column of a Green's function independently.
Naturally, they each contain information about the environment that the other calculations might use for embedding.
Using this information would effectively exchange an increased communication cost for increased accuracy.
This same rationale applied to hybrid algorithms where we want to extract global information
 from randomized calculations to improve localized calculations.

We articulate the Green's function embedding problem in a simple notation.
One column of a Green's function matrix is written as the solution to a block linear system of the form
\begin{equation} \label{local_partition}
 \begin{bmatrix} \mathbf{H}_{LL} & \mathbf{H}_{LE} \\ \mathbf{H}_{EL} & \mathbf{H}_{EE} \end{bmatrix} \begin{bmatrix} \mathbf{g}_L \\ \mathbf{g}_E \end{bmatrix}
 = \begin{bmatrix} \mathbf{e}_1 \\ 0 \end{bmatrix},
\end{equation}
 for an implicitly complex-shifted Hamiltonian matrix that has been appropriately permuted and partitioned.
An embedding self-energy $\mathbf{\Sigma}_{\mathrm{embed}}$ can be defined to reduce the system to the ``local'' block of the linear system,
 which is equivalent to the Schur complement of the ``environment'' block \cite{PEXSI_Sigma},
\begin{subequations}
\begin{align} \label{local_reduced}
 \left( \mathbf{H}_{LL} + \mathbf{\Sigma}_{\mathrm{embed}} \right) \mathbf{g}_L &= \mathbf{e}_1, \\
 \mathbf{\Sigma}_{\mathrm{embed}} &= -\mathbf{H}_{LE}  \mathbf{H}_{EE}^{-1}  \mathbf{H}_{EL}.
\end{align}
\end{subequations}
Localized calculations assume the approximation $\mathbf{\Sigma}_{\mathrm{embed}} \approx 0$.
Non-trivial approximations for $\mathbf{\Sigma}_{\mathrm{embed}}$ can be mathematical or physical.
An example mathematical approximation is solving the localized problem in a least-squares sense,
\begin{equation}
 \min_{\mathbf{g}_L} \left\| \begin{bmatrix} \mathbf{H}_{LL} \\ \mathbf{H}_{EL} \end{bmatrix} \mathbf{g}_L - \begin{bmatrix} \mathbf{e}_1 \\ 0 \end{bmatrix} \right\|_2 ,
\end{equation}
which corresponding to the self-energy approximation
\begin{equation} \label{least_squares_self_energy}
 \mathbf{\Sigma}_{\mathrm{embed}} \approx \mathbf{H}_{LL}^{-\dag} \mathbf{H}_{EL}^\dag \mathbf{H}_{EL} .
\end{equation}
An example physical approximation is to use the self-energy of free electrons in the environment,
 which converges rapidly with localization radius for simple metals like copper \cite{local_scattering}.

The localization self-energy $\mathbf{\Sigma}_{\mathrm{local}}$ is a global concept that
 complements the local concept of an embedding self-energy.
We define it using the inverse, $\mathbf{G} \equiv \mathbf{H}^{-1}$, of the unpartitioned matrix $\mathbf{H}$ from Eq.\@ (\ref{local_partition})
  and a sparse approximation $\tilde{\mathbf{G}}$ as
\begin{equation}
 \mathbf{\Sigma}_{\mathrm{local}} \equiv  \mathbf{G}^{-1} - \tilde{\mathbf{G}}^{-1} .
\end{equation}
We cannot perform an exact inversion on these large matrices efficiently,
 and $\mathbf{\Sigma}_{\mathrm{local}}$ needs further structure or assumptions to be useful computationally.
If it is small in norm, then we can use a perturbation series to improve the accuracy of $\tilde{\mathbf{G}}$,
\begin{align}
  \mathbf{G} &= \tilde{\mathbf{G}} - \tilde{\mathbf{G}} \mathbf{\Sigma}_{\mathrm{local}} \tilde{\mathbf{G}} 
 + \tilde{\mathbf{G}}  \mathbf{\Sigma}_{\mathrm{local}} \tilde{\mathbf{G}} \mathbf{\Sigma}_{\mathrm{local}}\tilde{\mathbf{G}}
 + O\left( \|\mathbf{\Sigma}_{\mathrm{local}} \|^3 \right) \notag \\
 &= 3 \tilde{\mathbf{G}} - 3 \tilde{\mathbf{G}}\mathbf{H}\tilde{\mathbf{G}} + \tilde{\mathbf{G}}\mathbf{H} \tilde{\mathbf{G}}\mathbf{H}\tilde{\mathbf{G}} + O\left( \|\mathbf{\Sigma}_{\mathrm{local}} \|^3 \right).
\end{align}
This could benefit selected inversion by partitioning a matrix, $\mathbf{H}=\tilde{\mathbf{H}} +  \mathbf{\Sigma}_{\mathrm{local}}$,
 into a more sparse $\tilde{\mathbf{H}}$ and a small-norm $\mathbf{\Sigma}_{\mathrm{local}}$
 and perturbatively correcting $\tilde{\mathbf{G}} = \tilde{\mathbf{H}}^{-1}$.
If it can be factored into a reduced-rank form, $\mathbf{\Sigma}_{\mathrm{local}} = \mathbf{X} \mathbf{\Sigma}_1 \mathbf{X}^{\dag}$,
 then we can apply the Woodbury formula for a reduced-rank update of $\tilde{\mathbf{G}}$ to $\mathbf{G}$,
\begin{subequations}
\begin{align}
 \mathbf{\Sigma}_2^{-1} &\equiv \mathbf{\Sigma}_1^{-1} + \mathbf{X}^{\dag} \tilde{\mathbf{G}} \mathbf{X}, \\
 \mathbf{G} &= \tilde{\mathbf{G}} - \tilde{\mathbf{G}} \mathbf{X} \mathbf{\Sigma}_2 \mathbf{X}^{\dag} \tilde{\mathbf{G}} .
\end{align}
\end{subequations}
In both examples, we have identified a beneficial structure for $\mathbf{\Sigma}_{\mathrm{local}}$
 that we can try to impose through our choice of $\tilde{\mathbf{G}}$.

We can use Newton's method for matrix inversion to relate $\mathbf{\Sigma}_{\mathrm{embed}}$ and $\mathbf{\Sigma}_{\mathrm{local}}$
 and conceptually fix one of the performance problems in the localized linear-scaling algorithm.
Newton's method generates a sequence of approximate inverses $\mathbf{G}_i$ by
\begin{equation}
 \mathbf{G}_{i+1} = 2 \mathbf{G}_i -  \mathbf{G}_i \mathbf{H} \mathbf{G}_i
\end{equation}
 that is quadratically convergent from an initial approximation $\mathbf{G}_0$.
Relative to using iterative linear solvers to compute each column of $\mathbf{G}$ independently
 with a linear convergence rate that depends on condition number,
 this process is effectively self-preconditioning.
However, truncations of $\mathbf{G}$ to $\tilde{\mathbf{G}}$ during this process
 is incompatible with quadratic convergence, just as in the related process of density-matrix purification \cite{truncation_stagnation}.
We can more robustly achieve the same self-preconditioning effect by adding nonlinearities into the equations defining $\tilde{\mathbf{G}}$ such as
\begin{subequations}
\begin{align}
 & \min_{\tilde{\mathbf{G}} = \mathbf{M} \odot \tilde{\mathbf{G}} }  \left\| \tilde{\mathbf{G}} -  \tilde{\mathbf{G}} \mathbf{H} \tilde{\mathbf{G}} \right\|_F   \ \ \mathrm{or} \\
 & \ \ \ \tilde{\mathbf{G}} = \mathbf{M} \odot \left( \tilde{\mathbf{G}} \mathbf{H} \tilde{\mathbf{G}} \right) . \label{self_energy_equation}
\end{align}
\end{subequations}
These both impose constraints on $\mathbf{\Sigma}_{\mathrm{local}}$ through its first-order perturbative correction to $\tilde{\mathbf{G}}$,
 to minimize the Frobenius norm or zero it within the sparsity pattern of $\tilde{\mathbf{G}}$.
In the latter case, the first column of $\tilde{\mathbf{G}}$ is equivalent to solving Eq.\@ (\ref{local_reduced}) for an
 approximate $\mathbf{\Sigma}_{\mathrm{embed}}$ constructed from partitioned blocks of $\tilde{\mathbf{G}}$,
\begin{equation} \label{optimized_self_energy}
  \mathbf{\Sigma}_{\mathrm{embed}} \approx \tilde{\mathbf{G}}_{LL}^{-1} \tilde{\mathbf{G}}_{LE} \mathbf{H}_{EL},
\end{equation}
 which is exact when $\tilde{\mathbf{G}}_{LE} = \mathbf{G}_{LE}$ and $\tilde{\mathbf{G}}_{LL} = \mathbf{G}_{LL}$.
This result satisfies the goal of having a nontrivial $\mathbf{\Sigma}_{\mathrm{embed}}$ approximation
 from coupling the independent-column calculations of $\tilde{\mathbf{G}}$.

\begin{figure}[t!]
\includegraphics{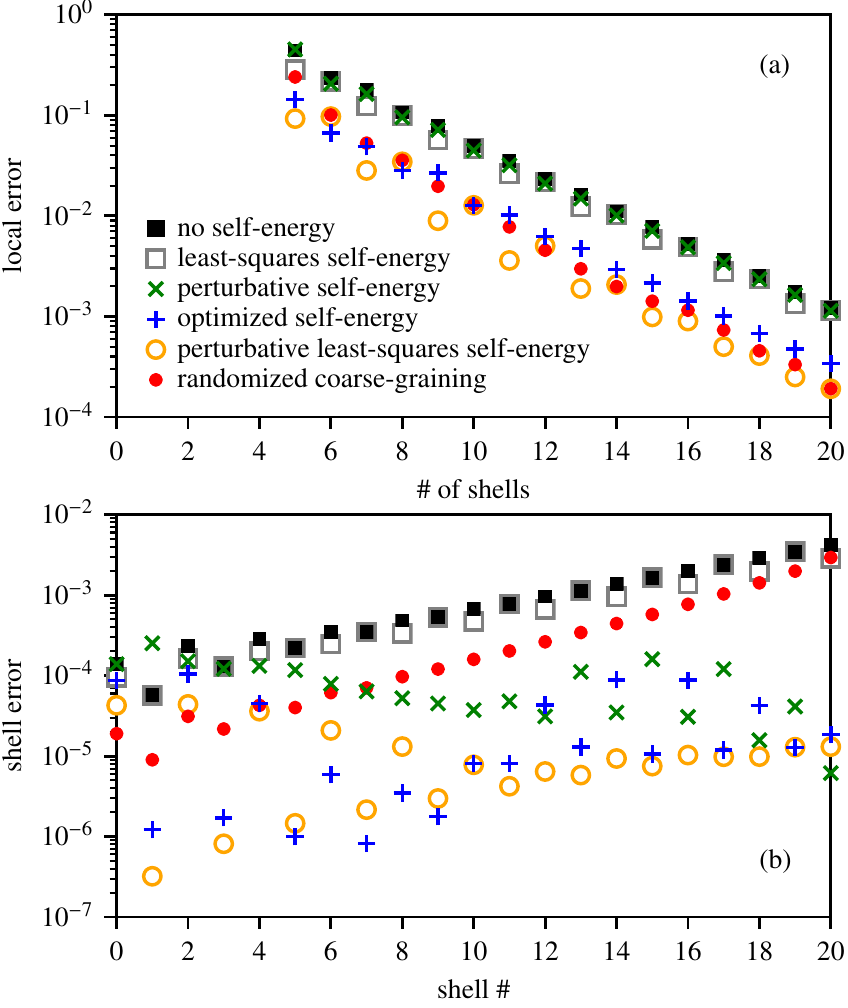}
\caption{\label{test_fig} Numerical tests of localization self-energy and randomized coarse-graining on the simple-cubic tight-binding model.
Frobenius norm error per site of a Green's function that is imaginary-shifted by the hopping energy is shown,
 (a) restricted to the first five shells for varying localization radii and
 (b) resolved by shell for a localization radius of 21 shells.
We compare conventional localized calculations with four distinct approximations of the localization self-energy
 and one implementation of randomized coarse-graining.
}
\end{figure}

We compare four approximations of the localization self-energy in Fig.\@ \ref{test_fig} on the simple-cubic tight-binding model.
The sites in a calculation are arranged into shells that are defined by the number of hops from a central site.
The least-squares self-energy approximation in Eq.\@ (\ref{least_squares_self_energy}) utilizes no information
 from other columns of the Green's function and does little to increase accuracy.
The first-order perturbative correction to the localized Green's function reduces errors in the outermost shells but does not reduce local error much.
When we then optimize the self-energy as in Eq.\@ (\ref{optimized_self_energy}) by solving Eq.\@ (\ref{self_energy_equation}),
 the error reduction spreads to the innermost shells to reduce local error.
Thus, utilizing information from other columns of the Green's function can improve accuracy
 but does not yet increase the rate of convergence with localization radius.
It is also not yet clear if the accuracy gained by solving Eq.\@ (\ref{self_energy_equation})
 justifies the increase in cost relative to solving Eq.\@ (\ref{local_reduced}).
We can obtain a comparable result more efficiently by applying a first-order perturbative correction to the Green's function with a least-squares self-energy,
 which have surprising synergy.

\subsection{Randomized coarse-graining}

The success of renormalization-group methods in statistical physics has motivated
 the development of multilevel methods in electronic structure and numerical linear algebra,
 but their successes have been more limited.
There are successful two-level electronic structure methods \cite{ONETEP_metal}
 that construct a small basis of local orbitals within a larger, systematic basis
 so that matrices can be restricted to the smaller basis,
 which in effect focuses the problem onto a narrower energy interval.
There is a multilevel generalization of this approach using temperature to adjust the energy interval \cite{energy_renorm},
 but competitive performance has not been demonstrated for greater than two levels.
Of the many types of multilevel methods in numerical linear algebra,
 structured matrix formats that are approximately closed under inversion are most relevant to electronic structure.
There are hierarchical matrix formats with this property \cite{hierarchical_matrices},
 but they are not yet general enough for the Hamiltonian matrices used in electronic structure.
They work well for smooth operators such as the Coulomb kernel and
 have been generalized to the free-electron Green's function \cite{directional_H2},
 but these concepts do not yet extend to more general sparse matrices.
An elegant high-accuracy solution to this problem might exist,
 but there is not yet a clear path to it.
Instead, we can initiate the development of low-accuracy methods using randomized algorithms.

The basic idea of randomized coarse-graining is to control the rank of off-diagonal matrix blocks
 by approximating them with randomized resolutions of identity as in Eq.\@ (\ref{random_RoI}).
We use this control to induce computationally useful matrix structure.
For example, when Eq.\@ (\ref{local_partition}) corresponds to a localized basis set in $D$ spatial dimensions,
 the ``environment'' matrix blocks can be further partitioned and permuted into ``shell'' blocks,
\begin{equation} \label{shell_partition}
 \begin{bmatrix} \mathbf{H}_{LL} & \mathbf{H}_{L1} & 0 & 0 & \cdots\vphantom{\mathbf{H}^1} \\ \mathbf{H}_{1L} & \mathbf{H}_{11} & \mathbf{H}_{12} & 0 & \cdots \\
 0 & \mathbf{H}_{21} & \mathbf{H}_{22} & \mathbf{H}_{23} & \cdots \\ 0 & 0 & \mathbf{H}_{32} & \mathbf{H}_{33} & \cdots \\
 \vdots & \vdots & \vdots & \vdots & \ddots
  \end{bmatrix} \begin{bmatrix} \mathbf{g}_L \\ \mathbf{g}_1 \\ \mathbf{g}_2 \\ \mathbf{g}_3 \\ \vdots \end{bmatrix}
 = \begin{bmatrix} \mathbf{e}_1 \\ 0 \\ 0 \\ 0 \\ \vdots \end{bmatrix},
\end{equation}
 where the matrix dimension of shell $s$ is $O(s^{D-1})$.
For $D=1$, this is a banded matrix with a bandwidth that does not depend on the number of shells.
For $D > 1$, the bandwidth grows with the number of shells.
We can control the matrix bandwidth if we approximate the off-diagonal blocks as
\begin{subequations}
\begin{align}
 \mathbf{H}_{L1} &\approx \overline{\mathbf{H}}_{L1} \mathbf{R}_1^\dag \ \ \ \mathrm{for} \ \ \  \overline{\mathbf{H}}_{L1} = \mathbf{H}_{L1} \mathbf{R}_1  , \\
 \mathbf{H}_{i(i+1)} &\approx \mathbf{R}_i \overline{\mathbf{H}}_{i(i+1)} \mathbf{R}_{i+1}^\dag \ \ \ \mathrm{for} \ \ \ 
 \overline{\mathbf{H}}_{i(i+1)} = \mathbf{R}_i^\dag \mathbf{H}_{i(i+1)} \mathbf{R}_{i+1} ,
\end{align}
\end{subequations}
 where $\mathbf{R}_i \mathbf{R}_i^\dag$ are random projectors that all have the same rank.
Enabled by this approximation, we coarsen the linear system to a block-tridiagonal form with smaller blocks,
\begin{equation} \label{shell_restriction}
 \begin{bmatrix} \mathbf{H}_{LL} & \overline{\mathbf{H}}_{L1} & 0 & 0 & \cdots\vphantom{\mathbf{H}^1} \\ \overline{\mathbf{H}}_{1L} & \overline{\mathbf{H}}_{11} &  \overline{\mathbf{H}}_{12} & 0 & \cdots \\
 0 &  \overline{\mathbf{H}}_{21}  & \overline{\mathbf{H}}_{22} & \overline{\mathbf{H}}_{23} & \cdots \\ 0 & 0 & \overline{\mathbf{H}}_{32} & \overline{\mathbf{H}}_{33} & \cdots \\
 \vdots & \vdots & \vdots & \vdots & \ddots
  \end{bmatrix} \begin{bmatrix} \mathbf{g}_L \\ \mathbf{R}_1^\dag \mathbf{g}_1 \\ \mathbf{R}_2^\dag \mathbf{g}_2 \\ \mathbf{R}_3^\dag \mathbf{g}_3 \\ \vdots \end{bmatrix}
 = \begin{bmatrix} \mathbf{e}_1 \\ 0 \\ 0 \\ 0 \\ \vdots \end{bmatrix},
\end{equation}
 where $\mathbf{H}_{ii}$ are deflated to their Schur complement 
 in the null space of $\mathbf{R}_i$ spanned by $\mathbf{U}_i$ with orthonormal columns,
\begin{subequations}
\begin{align}
 \mathbf{\Sigma}_i &\equiv  - \mathbf{H}_{ii} \mathbf{U}_i (\mathbf{U}_i^\dag \mathbf{H}_{ii} \mathbf{U}_i)^{-1} \mathbf{U}_i^\dag \mathbf{H}_{ii} , \\
 \overline{\mathbf{H}}_{ii} &= (\mathbf{R}_i^\dag \mathbf{R}_i)^{-1} \mathbf{R}_i^\dag ( \mathbf{H}_{ii} +  \mathbf{\Sigma}_i ) \mathbf{R}_i (\mathbf{R}_i^\dag \mathbf{R}_i)^{-1}.
\end{align}
\end{subequations}
This is a representative example but not a practical algorithm.
The deflation of $\mathbf{H}_{ii}$ is not efficient, and it is difficult to reuse this coarse-graining procedure for other source vectors
 since they will have different shell partitionings.
Practical coarse-graining needs hierarchical structure to fix these problems.

We test the shell-based coarse-graining of Eq.\@ (\ref{shell_restriction}) on the simple-cubic tight-binding model in Fig.\@ \ref{test_fig}.
With respect to a target shell, we retain the inner shells exactly and coarsen the outer shells down to the dimension of the target shell.
We see a consistent reduction of errors with the largest reductions for the shells farthest from the coarsened region.
The local error reduction is comparable to the best self-energy approximation that we have tested,
 but there are larger outer-shell errors.

For this application, random low-rank projectors are meant for approximating traces with the coarsened matrices and are
 not intended to produce accurate low-rank approximations of the off-diagonal matrix blocks that they project.
The generic form of these approximations within a trace calculation is
\begin{equation}
 \mathrm{tr}\left[ \mathbf{X} \right] \approx \mathrm{tr}\left[ \mathbf{X} \mathbf{R} \mathbf{R}^\dag \right] ,
\end{equation}
 where $\mathbf{X}$ is unrelated to the off-diagonal block being projected
 and $\mathbf{R}$ has $m$ rows and $r$ columns.
If we use Gaussian random vectors where $[\mathbf{E}]_{i,j} = 1$, then Eq.\@ (\ref{sampling_error}) reduces to
\begin{equation}\label{randomproj_error}
 \left|\mathrm{tr}\left[ \mathbf{X} - \mathbf{X} \mathbf{R} \mathbf{R}^\dag \right] \right| \approx \frac{\| \mathbf{X} \|_F}{\sqrt{r}} .
\end{equation}
By contrast, if $\mathbf{R} \mathbf{R}^\dag$ is a strict projector that creates an optimal low-rank approximation
 of an off-diagonal matrix block with full rank, then the comparable error estimate is
\begin{equation} \label{lowrank_error}
 \left|\mathrm{tr}\left[ \mathbf{X} - \mathbf{X} \mathbf{R} \mathbf{R}^\dag \right] \right| \approx \| \mathbf{X} \|_F \left( 1 - \frac{r}{m} \right) .
\end{equation}
These two types of projectors critically differ in how they are normalized.
The projectors for low-rank approximation have orthonormal columns, $\mathbf{R}^\dag \mathbf{R} = \mathbf{I}$,
 while the random projectors are normalized for trace preservation, $\mathbf{R}^\dag \mathbf{R} \approx (m/r) \mathbf{I}$.
For an error target $\epsilon$, Eq.\@ (\ref{randomproj_error}) has a $r \propto \epsilon^{-1/2}$ scaling while Eq.\@ (\ref{lowrank_error})
 requires $r \approx m$, which is not useful for coarse graining.

To develop a theoretical framework for randomized coarse-graining,
 one possible approach is to generalize the concepts of Sec.\@ \ref{trace_approx}
 into randomized, vector-dependent resolutions of identity.
We can try to rationalize the example in Eq.\@ (\ref{shell_restriction}) as
\begin{subequations} \label{vdroi1}
\begin{align}
 \mathbf{R} &= \begin{bmatrix} \mathbf{I} & 0 & 0 & 0 & \cdots\vphantom{\mathbf{H}^1} \\ 0 & \mathbf{R}_1 & 0 & 0 & \cdots \\ 
 0 & 0 & \mathbf{R}_2 & 0 & \cdots \\ 0 & 0 & 0 & \mathbf{R}_3 & \cdots \\ \vdots & \vdots & \vdots & \vdots & \ddots \end{bmatrix}, \\
 \mathbf{R}^\dag \mathbf{g} &\approx (\mathbf{R}^\dag \mathbf{H} \mathbf{R})^{-1} \mathbf{R}^\dag \mathbf{e}_1.
\end{align}
\end{subequations}
However, this is not equivalent to Eq.\@ (\ref{shell_restriction}) and hides a subtle normalization problem for the diagonal shell blocks.
Matrix inversion can be interpretted as a matrix polynomial, which is nonlinear in the resolution of identity.
Whenever two random resolutions of identity occur in a product, they introduce bias if they are the same random instance.
A worst-case example is $\mathrm{tr}[\mathbf{R}_1 \mathbf{R}_1^\dag \mathbf{R}_1 \mathbf{R}_1^\dag] \approx m^2 / r$
 versus $\mathrm{tr}[\mathbf{R}_1 \mathbf{R}_1^\dag \mathbf{R}_2 \mathbf{R}_2^\dag] \approx m$
 for a pair of $m$-by-$r$ random instances $\mathbf{R}_1$ and $\mathbf{R}_2$.
Whether or not bias can be removed efficiently or at least controllably reduced is an important
 open problem in developing this framework.

When viewed as a matrix approximation, vector-dependent resolutions of identity are linear operators on matrices.
They decompose into outer products acting on the left or right of a matrix,
 but the left and right products are not independent.
A symmetric example of this matrix-approximation form is
\begin{equation} \label{vdroi2}
 \tilde{\mathbf{X}} = \sum_{(i,j) \in \mathcal{P}} \mathbf{r}_i \mathbf{r}_i^\dag \mathbf{X} \mathbf{r}_j \mathbf{r}_j^\dag ,
\end{equation}
 where $\mathcal{P}$ is the admissible set of vector pairs.
While Eq.\@ (\ref{vdroi1}) envisions a restricted matrix inversion for approximating one column of a matrix inverse,
 we can also use Eq.\@ (\ref{vdroi2}) to define a more monolithic approximate inverse $\tilde{\mathbf{G}}$ of a matrix $\mathbf{H}$ as
\begin{equation}
 \sum_{(i,j),(j,k),(i,k) \in \mathcal{P}} \mathbf{r}_i \mathbf{r}_i^\dag \mathbf{H} \mathbf{r}_j \mathbf{r}_j^\dag \tilde{\mathbf{G}} \mathbf{r}_k \mathbf{r}_k^\dag \approx \sum_{(i,j) \in \mathcal{P}} \mathbf{r}_i \mathbf{r}_i^\dag \mathbf{r}_j \mathbf{r}_j^\dag.
\end{equation}
Thus there are many possible mathematical interpretations of the physical concept of randomized coarse-graining,
 and each can produce a different numerical method with a different cost and accuracy.
More research is needed to explore these ideas and identify the most promising numerical methods.

\section{Conclusions\label{conclusions}}\vspace*{-0.3cm}

This study of linear-scaling electronic structure algorithms has been limited in scope
 to a simple set of model systems, an unoptimized software implementation, and modest computing resources,
 but we can still conclude a lot from it.
Benchmarks in Fig.\@ \ref{scaling_fig} for metals at $T \ge 1$ eV are comparable to previous benchmarks for insulators \cite{TB_bench},
 but simulation costs rapidly grow as the temperature is decreased towards ambient.
We are unable to find a regime where randomized algorithms improve upon the performance of localized algorithms.
The crossover point for linear-scaling algorithms is delayed to larger system sizes
 because subcubic-scaling selected inversion (PEXSI) is favorable in intermediate size regimes.
Thus the competitive regime for linear-scaling algorithms has only become smaller since their active research period in the 1990's \cite{linear_quantum},
 and low-$T$ metals remain a particular challenge \cite{linear_metal_review}.
A broader scope of study is unlikely to change these conclusions: larger basis sets
 further disfavor linear-scaling algorithms, optimized software can only reduce our linear-scaling cost prefactors
 by a factor of eight before we saturate the mature performance of dense linear algebra as noted in Sec.\@ \ref{performance},
 and more heterogeneous, larger-scale computing resources do not fundamentally favor any of
 the assessed algorithms while significantly increasing the effort required for efficient software implementations.

While our assessment of available linear-scaling algorithms is rather negative, there are some positive aspects to it.
In the process of coming to these conclusions, we have developed a unified theoretical framework
 that can describe both localized and randomized algorithms.
We have used it to explore two promising theoretical concepts in Sec.\@ \ref{discussion}
 that might develop into successful hybridized linear-scaling algorithms.
Also, the use of randomized algorithms to unbias force errors and drive Langevin dynamics might still be useful
 even if it adds to the cost of localized calculations and small biases remain.

Rational approximations of important matrix functions for electronic structure are almost ready for general use,
 and our benchmarks show that at low $T$ they out-perform polynomial approximations even without effective preconditioners.
What remains to be done are optimizing rational approximations of the free-energy function $g(x)$ in Eq.\@ (\ref{free_energy_function})
 to use the analytical free-energy derivatives in Sec.\@ \ref{analytic_derivative}
 and a more efficient tuning procedure for the chemical potential \cite{rational_tuning}.
Any improvements to preconditioners or localization self-energy approximations can further enhance their performance.
The rational form also enables finite-$T$ generalizations of density-matrix purification through the use of Newton's method for matrix inversion.
We should also consider the switch from iterative to direct linear solvers for local calculations
 with efficient updates of matrix factorizations when local matrices only differ by the insertion and removal of a small number of rows and columns.

The persistent challenges in general-purpose linear-scaling electronic structure require new algorithms
 for low-accuracy but high-reliability numerical linear algebra.
Most research in numerical linear algebra research is focused on high-accuracy algorithms
 with a notable exception of preconditioners.
They tolerate a wide range of accuracy since iterative linear solvers naturally repair many of their shortcomings.
This is no longer the case when we want a structured approximate inverse and not just a preconditioner.
Randomized algorithms are used in numerical linear algebra, notably for high-accuracy low-rank approximation \cite{random_rank}.
Their use for low-accuracy applications such as preconditioning is more recent and ongoing \cite{STRUMPACK},
 but this is a promising area for randomized algorithms because of the inherently poor dependence of their cost on accuracy.
If a structured approximate inverse algorithm can achieve the low accuracy targets in this paper at a competitive cost,
 then it can have a large practical impact on electronic structure.
Analysis of low-accuracy algorithms may benefit from the development and use of error estimates based on statistical assumptions
 in place of strict error bounds, which are often not tight enough to be directly useful in the low-accuracy regime.

It would also be worthwhile to develop new semiempirical electronic structure models to fill the gap between interatomic potentials and first-principles electronic structure.
While the development of DFTB is still active, semiempirical quantum chemistry has only a few remaining developers \cite{PM7,OM3},
 and planewave-based semiempirical electronic structure \cite{uniform_semiempirical} has never developed total-energy models
 even though it has good computational performance \cite{large_semiempirical}.
Modernized semiempirical electronic structure models could incorporate model-building concepts from machine learning
 and relevant ideas from the last several decades of method development in first-principles electronic structure.
With improved accuracy and reliability, semiempirical models could be viable for more applications.
Semiempiricism also enables a unique opportunity to operate linear-scaling solver algorithms at fixed cost rather than fixed accuracy.
Uncontrolled solver errors can be minimized on an equal footing with finite-basis and electron-correlation errors
 during the parameterization of a semiempirical model.
Such a model would be associated with a specific solver algorithm, in contrast to the standard practice
 of numerical interoperability between solver algorithms as in the ELSI project \cite{ELSI}.

Ultimately, superlinear costs have a strong impact on how electronic structure simulations are applied.
Cost reductions drive applications towards the smallest acceptable number of atoms per simulation
 for converging or at least understanding finite-size effects.
These considerations favor time averaging and sample averaging over spatial averaging.
In applications to point defects, line defects, and surfaces, it is likely that the convergence of finite-size effects
 will always occur before the crossover point of linear-scaling algorithms.
In more general applications with long-range and large-scale electronic effects in highly inhomogeneous systems,
 it can be difficult to extract useful information from small simulations.
This is where we need a viable linear-scaling electronic structure capability.

\begin{acknowledgments}\vspace*{-0.3cm}
We thank Luke Shulenburger and Kipton Barros for useful discussions.
This work was supported by the Advanced Simulation and Computing for Physics and Engineering Models program at Sandia National Laboratories.
Sandia National Laboratories is a multimission laboratory managed and operated by National Technology and Engineering Solutions of Sandia, LLC.,
 a wholly owned subsidiary of Honeywell International, Inc., for the U.S. Department of Energy's National Nuclear Security Administration under contract DE-NA-0003525.
This paper describes objective technical results and analysis. Any subjective views or opinions that might be expressed in the paper
 do not necessarily represent the views of the U.S. Department of Energy or the United States Government.
\end{acknowledgments}

\end{document}